\documentclass[a4paper,UKenglish,cleveref, autoref, thm-restate]{lipics-v2021}

\usepackage[numbers,sort]{natbib}
\usepackage{graphicx}
\usepackage{latexsym,amsmath,textcomp,pifont,marvosym,wasysym,amssymb}

\usepackage[algo2e,ruled,vlined,linesnumbered, noend]{algorithm2e}
\usepackage[utf8]{inputenc}
\usepackage[TS1,T1]{fontenc}
\SetCommentSty{textit}
\DontPrintSemicolon
\SetAlCapHSkip{0pt}
\SetAlgoLined

\SetKwFor{ParallelFor}{for}{do in parallel}{}
\SetKwFor{ParallelWhile}{while}{do in parallel}{}
\SetKwIF{If}{ElseIf}{Else}{if}{}{else if}{else}{end if}

\usepackage{placeins}
\usepackage{tikz}
\usetikzlibrary{calc}

\usepackage{soul}
\usepackage{multirow}
\usepackage{bm}
\usepackage{datatool}
\usepackage{booktabs}
\DTLsetseparator{ = }

\usepackage{pgfplots}
\usepgfplotslibrary{external}
\newif\ifpdfplots
\pdfplotstrue

\usepackage{rotating}

\usepackage{mathtools}

\newif\ifenablecomments
\enablecommentstrue

\newcommand{\placeholder}[2]{\DTLfetch{#1}{key}{#2}{value}}

\newcommand{\etal}{{et al}.\@ }
\newcommand{\myparagraph}[1]{\vspace{-0.35cm} \subparagraph*{#1}}

\newcommand{\mtkahypar}{\texttt{Mt-KaHyPar}}

\newcommand{\mtkahyparold}{Mt-KaHyPar-D}

\newcommand{\Oh}[1]{\ensuremath{\mathcal{O}(#1)}}

\newcommand{\overbar}[1]{\mkern 1.5mu\overline{\mkern-1.5mu#1\mkern-1.5mu}\mkern 1.5mu}

\newcommand{\Partition}{\ensuremath{\mathrm{\Pi}}}%
\newcommand{\incnets}{\ensuremath{\mathrm{I}}}%
\newcommand{\pinsinpart}{\ensuremath{\mathrm{\Phi}}}

\newcommand{\con}{\ensuremath{\lambda}}
\newcommand{\conset}{\ensuremath{\Lambda}}

\newcommand{\maxsize}[1]{\ensuremath{\Delta_{#1}}}

\newcommand{\avgsize}{\ensuremath{\overline{|e|}}}
\newcommand{\meddeg}{\ensuremath{\widetilde{d(v)}}}
\newcommand{\medsize}{\ensuremath{\widetilde{|e|}}}

\newcommand{\quotientgraph}{\ensuremath{\mathcal{Q}}}
\newcommand{\activequeue}{\ensuremath{Q}}

\newcommand{\Partitioner}[1]{\textsf{#1}} 

\newcommand{\flownetwork}{\ensuremath{\mathcal{N}}}
\newcommand{\flowhypergraph}{\ensuremath{\mathcal{H}}}
\newcommand{\flownodeset}{\ensuremath{\mathcal{V}}}
\newcommand{\flowedgeset}{\ensuremath{\mathcal{E}}}
\DeclareMathAlphabet{\mathpzc}{OT1}{pzc}{m}{n}
\newcommand{\capacity}{\ensuremath{\mathpzc{c}}}
\newcommand{\source}{\ensuremath{s}}
\newcommand{\sink}{\ensuremath{t}}
\newcommand{\sourcesink}{\ensuremath{(\source,\sink)}}
\newcommand{\innode}{\ensuremath{e_{\text{in}}}}
\newcommand{\outnode}{\ensuremath{e_{\text{out}}}}
\newcommand{\distance}{\ensuremath{d}}
\newcommand{\excess}{\ensuremath{\operatorname{exc}}}

\newcommand{\mtkahypardflows}{\Partitioner{Mt-KaHyPar-D-F}}
\newcommand{\mtkahyparqflows}{\Partitioner{Mt-KaHyPar-Q-F}}
\newcommand{\mtkahypardflowsconfig}[1]{\mtkahypardflows~\Partitioner{#1}}
\newcommand{\mtkahyparqflowsconfig}[1]{\mtkahyparqflows~\Partitioner{#1}}

\definecolor{fuchsiapink}{rgb}{1.0, 0.47, 1.0}
\definecolor{amethyst}{rgb}{0.6, 0.4, 0.8}

\definecolor{utahcrimson}{rgb}{0.83, 0.0, 0.25}

\definecolor{ao}{rgb}{0.0, 0.5, 0.0}

\newcommand{\plusplus}{\texttt{++}}

\newcommand{\gpp}[1]{g\plusplus#1}

\newcommand{\splitatcommas}[1]{%
  \begingroup
  \begingroup\lccode`~=`, \lowercase{\endgroup
    \edef~{\mathchar\the\mathcode`, \penalty0 \noexpand\hspace{0pt plus 1em}}%
  }\mathcode`,="8000 #1%
  \endgroup
}

\title{Parallel Flow-Based Hypergraph Partitioning} 

\titlerunning{Parallel Flow-Based Hypergraph Partitioning} 

\author{Lars Gottesbüren}{Karlsruhe Institute of Technology, Karlsruhe, Germany}{lars.gottesbueren@kit.edu}{}{}

\author{Tobias Heuer}{Karlsruhe Institute of Technology, Karlsruhe, Germany}{tobias.heuer@student.kit.edu}{}{}

\author{Peter Sanders}{Karlsruhe Institute of Technology, Karlsruhe, Germany}{sanders@kit.edu}{}{}

\authorrunning{L. Gottesbüren, T. Heuer and P. Sanders} 

\Copyright{Lars Gottesbüren, Tobias Heuer and Peter Sanders} 



\relatedversion{} 

\supplement{
	\\
	Source code (multilevel framework): \url{https://github.com/kahypar/mt-kahypar} \\
	Source code (flow-based refinement): \url{https://github.com/larsgottesbueren/WHFC/tree/parallel}
	\\ Benchmark Set \& Experimental Results: \url{https://algo2.iti.kit.edu/heuer/sea22/}
}

\funding{This work was partially supported by DFG grants WA654/19-2 and SA933/11-1. The authors acknowledge support by the state of
Baden-Württemberg through bwHPC.}

\nolinenumbers 

\hideLIPIcs  

\EventEditors{John Q. Open and Joan R. Access}
\EventNoEds{2}
\EventLongTitle{42nd Conference on Very Important Topics (CVIT 2016)}
\EventShortTitle{CVIT 2016}
\EventAcronym{CVIT}
\EventYear{2016}
\EventDate{December 24--27, 2016}
\EventLocation{Little Whinging, United Kingdom}
\EventLogo{}
\SeriesVolume{42}
\ArticleNo{23}

\begin{document}

\maketitle

\begin{abstract}
We present a shared-memory parallelization of \emph{flow-based refinement}, which is considered the most powerful iterative improvement technique for hypergraph partitioning at the moment.
Flow-based refinement works on bipartitions, so current sequential partitioners schedule it on different block pairs to improve $k$-way partitions.
We investigate two different sources of parallelism: a parallel scheduling scheme and a parallel maximum flow algorithm based on the well-known push-relabel algorithm.
In addition to thoroughly engineered implementations, we propose several optimizations that substantially accelerate the algorithm in practice, enabling the use on extremely large hypergraphs (up to $1$ billion pins).
We integrate our approach in the state-of-the-art parallel multilevel framework \Partitioner{Mt-KaHyPar} and conduct extensive experiments on a benchmark set of more
than 500 real-world hypergraphs, to show that the partition quality of our code is on par with the highest quality sequential code (\Partitioner{KaHyPar}), while being
an order of magnitude faster with 10 threads.

\end{abstract}

\section{Introduction}
Balanced hypergraph partitioning is a classical NP-hard optimization problem with numerous applications.
Hypergraphs are a generalization of graphs, where each hyperedge can connect an arbitrary number of vertices.
The problem is to partition the vertices of a hypergraph $H = (V,E,\omega)$ into $k$ disjoint blocks $V_1, \dots V_k$ of roughly equal size
($\forall V_i: |V_i| \le (1+\varepsilon)\frac{|V|}{k}$), such that an objective function defined on the hyperedges is minimized.
In this work, we consider the connectivity metric $\sum_{e \in E} (\lambda(e) - 1) \omega(e)$ where $\lambda(e) := \{ V_i \mid e \cap V_i \neq \emptyset\}$
denotes the number of different blocks connected by hyperedge $e \in E$ and $\omega(e)$ denotes its weight.
Often balanced partitioning is used as an acceleration technique for other applications, such as quantum circuit simulation~\cite{gray2021hyper},
sharding distributed databases~\cite{schism, SHP}, load balancing (for scientific computing)~\cite{PATOH}, route planning~\cite{delling2010graph, FLOW-CUTTER}, or boosting cache utilization in a search engine backend~\cite{google-search}.

There is a substantial amount of literature on graph and hypergraph partitioning, which is why we refer to survey articles~\cite{ALPERT-SURVEY,GRAPH-SURVEY,PAPA-MARKOV,KAHYPAR-DIS} for a summary.
Due to being NP-hard~\cite{LENGAUER} and hard to approximate within constant factors~\cite{BUI}, most of the work focuses on heuristic algorithms, with the multilevel paradigm emerging as the most successful and widely used approach in modern partitioners~\cite{MT-METIS, HMETIS, PATOH, ZOLTAN, KAFFPA, MT-KAHIP, KAHYPAR-DIS, MT-KAHYPAR}.
Most partitioners use variations of well-known local vertex moving heuristics such as Kernighan-Lin~\cite{KL} or Fiduccia-Mattheyses~\cite{FM}
However, these techniques are known to easily get stuck in local minima~\cite{DBLP:journals/tc/Krishnamurthy84}.

In this situation, maximum flows are an excellent tool as they correspond to (unbalanced) minimum cuts, thus offering a more global view than local routines.
While being a fundamental algorithmic tool, maximum flows were long overlooked for partitioning heuristics, due to their complexity~\cite{yang-wong-fbb},
but have since enjoyed wide-spread adoption~\cite{yang-wong-fbb, lang2004flow, andersen2008algorithm, delling2010graph, KAFFPA, FLOW-CUTTER} in many different algorithmic
contexts, even outside the multilevel framework.

\myparagraph{Contribution.}

In this paper, we parallelize \emph{flow-based refinement}, a powerful technique that is the last missing component in a series of works~\cite{MT-KAHYPAR, MT-KAHYPAR-Q} on parallelizing the state-of-the-art multilevel hypergraph partitioning framework \Partitioner{KaHyPar}~\cite{KAHYPAR-DIS}.
Flow-based refinement operates on bipartitions, or on two blocks at a time if used for $k > 2$.
Scheduling independent block pairs gives some trivial parallelism.
One contribution we make is to improve the parallelism in the scheduler by relaxing certain constraints of the original approach and showing how to deal with race conditions caused by the relaxation.
For small $k$ this is still insufficiently parallel, which is why we also parallelize the refinement on two blocks.
We adapt an existing parallel flow algorithm to handle the incremental flow problems of the \Partitioner{FlowCutter} refinement algorithm~\cite{FLOW-CUTTER, REBAHFC, KAHYPAR-HFC}.
Additionally, we engineer an efficient implementation, proposing several optimizations that reduce running time in practice, and fix a so far not documented bug in the parallel flow algorithm itself, not the implementation.
The result is a partitioning framework that achieves the same solution quality as the highest quality sequential framework, but in a fast parallel code.
Using 10 threads, our code is an order of magnitude faster than sequential \Partitioner{KaHyPar} with flow-based refinement.

\myparagraph{Outline.}

The paper is organized as follows.
In Section~\ref{sec:preliminaries} we introduce notation, terminology, and some algorithmic preliminaries.
Section~\ref{sec:related_work} briefly deals with related work.
More details on additional related work are given in the main sections~\ref{sec:scheduling} -~\ref{sec:parallel_flows}, closer to where particular parts are needed.
In Section~\ref{sec:overview}, we give an overview of the different components in the framework and how they interact.
We complement the algorithmic discussion with extensive experiments in Section~\ref{sec:experiments}, before concluding in Section~\ref{sec:conclusion}.

\section{Preliminaries}\label{sec:preliminaries}
\subparagraph*{Hypergraphs.}
A \emph{weighted hypergraph} $H=(V,E,c,\omega)$ is defined as a set of vertices $V$ and a set of hyperedges/nets $E$
with vertex weights $c:V \rightarrow \mathbb{R}_{>0}$ and net
weights $\omega:E \rightarrow \mathbb{R}_{>0}$, where each net $e$ is
a subset of the vertex set $V$.
The vertices of a net are called its \emph{pins}.
We extend $c$ and $\omega$ to sets in the natural way, i.e., $\capacity(U) :=\sum_{v\in U} \capacity(v)$ and $\omega(F) :=\sum_{e \in F} \omega(e)$.
A vertex $v$ is \emph{incident} to a net $e$ if $v \in e$.
$\mathrm{I}(v)$ denotes the set of all incident nets of $v$.
The \emph{degree} of a vertex $v$ is $d(v) := |\mathrm{I}(v)|$.
The \emph{size} $|e|$ of a net $e$ is the number of its pins.
We call two nets $e_i$ and $e_j$ \emph{identical} if $e_i = e_j$.
Given a subset $V' \subset V$, the \emph{subhypergraph}
$H_{V'}$ is defined as $H_{V'}:=(V', \{e \cap V' \mid e \in E : e \cap V' \neq \emptyset \})$.

\myparagraph{Balanced Hypergraph Partitioning.}
A \emph{$k$-way partition} of a hypergraph $H$ is a function $\Partition : V \to \{1, \dots, k\}$.
The blocks $V_i := \Partition^{-1}(i)$ of $\Partition$ are the inverse images.
We call $\Partition$ \emph{$\varepsilon$-balanced} if each block $V_i$ satisfies the \emph{balance constraint}:
$\capacity(V_i) \leq L_{\max} := (1+\varepsilon) \lceil \frac{\capacity(V)}{k} \rceil$ for some parameter $\mathrm{\varepsilon} \in (0,1)$. 
For each net $e$, $\conset(e) := \{V_i \mid  V_i \cap e \neq \emptyset\}$ denotes the \emph{connectivity set} of $e$.
The \emph{connectivity} $\con(e)$ of a net $e$ is $\con(e) := |\conset(e)|$.
A net is called a \emph{cut net} if $\con(e) > 1$.
A node $u$ that is incident to at least one cut net is called \emph{boundary node}.
The number of pins of a net $e$ in block $V_i$ is denoted by $\pinsinpart(e,V_i) := |e \cap V_i|$.
The \emph{quotient graph} $\quotientgraph := (\Partition, E_{\Partition} := \{(V_i, V_j)~|~\exists e \in E: \{V_i,V_j\} \subseteq \conset(e)\})$ contains an edge between each pair of adjacent blocks of a $k$-way partition $\Partition$.
Given parameters $\varepsilon$ and $k$, and a hypergraph $H$, the \emph{balanced hypergraph partitioning problem} is to find an $\varepsilon$-balanced $k$-way partition $\Partition$
that minimizes an objective function defined on the hyperedges.
In this work, we minimize the \emph{connectivity metric} $(\lambda - 1)(\Pi) := \sum_{e \in E} (\lambda(e) - 1) \: \omega(e)$.

\myparagraph{Flows.}
A flow network $\flownetwork = (\flownodeset,\flowedgeset,\capacity)$ is a directed graph with a dedicated source
$\source \in \flownodeset$ and sink $\sink \in \flownodeset$ in which
each edge $e \in \flowedgeset$ has capacity $\capacity(e) \ge 0$. An $\sourcesink$-flow is a function
$f: \flownodeset \times \flownodeset \rightarrow \mathbb{R}$
that satisfies the \emph{capacity constraint} $\forall u,v \in \flownodeset: f(u,v) \le \capacity(u,v)$,
the \emph{skew symmetry constraint} $\forall u,v \in \flownodeset: f(u,v) = -f(v,u)$ and the
\emph{flow conservation constraint} $\forall u \in \flownodeset \setminus \{\source,\sink\}: \sum_{v \in \flownodeset} f(u,v) = 0$.
The value of a flow $|f| := \sum_{v \in \flownodeset} f(\source,v) = \sum_{v \in \flownodeset} f(v,\sink)$ is defined as the
total amout of flow transferred from $\source$ to $\sink$.
An $\sourcesink$-flow $f$ is a maximum $\sourcesink$-flow if there exists no other $\sourcesink$-flow $f'$ with $|f| < |f'|$.
The \emph{residual capacity} is defined as $r_f(e) = \capacity(e) - f(e)$.
An edge $e$ is \emph{saturated} if $r_f(e) = 0$.
$\flownetwork_f = (\flownodeset, \flowedgeset_f, r_f)$ with
$\flowedgeset_f := \{(u,v) \in \flownodeset \times \flownodeset \mid r_f(u,v) > 0\}$ is the \emph{residual network}.
The max-flow min-cut theorem states that the value $|f|$ of a maximum $\sourcesink$-flow equals the weight
of a minimum cut that separates $\source$ and $\sink$~\cite{MINCUT-THEOREM}. This is also called a \emph{minimum $\sourcesink$-cut}.
The source-side cut can be computed by BFS from the source via residual edges, the sink-side cut via reverse BFS from the sink.

\myparagraph{Push-Relabel Algorithm.}
The \emph{push-relabel} \cite{PUSH-RELABEL} maximum flow algorithm stores a distance label $\distance(u)$ and an excess value $\excess(u) := \sum_{v \in \flownodeset} f(v,u)$ for each node.
It maintains a \emph{preflow}~\cite{PREFLOW} which is a flow where the conservation constraint is replaced by $\excess(u) \geq 0$.
A node $u \in \flownodeset$ is \emph{active} if $\excess(u) > 0$. An edge $(u,v) \in \flowedgeset$ is \emph{admissible}
if $r_f(u,v) > 0$ and $\distance(u) = \distance(v) + 1$.
A \emph{$\text{push}(u,v)$} operation sends $\delta = \min(\excess(u),r_f(u,v))$ flow units over $(u,v)$.
It is applicable if $u$ is active and $(u,v)$ is admissible.
A \emph{$\text{relabel}(u)$} operation updates the distance label of $u$ to $\min(\{\distance(v) + 1 \mid r_f(u,v) > 0\})$, which is applicable if $u$ is active has no admissible edges.
The distance labels are initialized to $\forall u \in \flownodeset - \source: \distance(u) = 0$ and $\distance(\source) = |V|$ and all source edges are saturated.
Efficient variants use the \emph{discharge} routine, which repeatedly scans the edges of an active node until its excess is zero.
All admissible edges are pushed and at the end of a scan, the node is relabeled.
Discharging active nodes in FIFO order results in an $\mathcal{O}(|\flownodeset|^3)$ time algorithm.
The \emph{global relabeling} heuristic~\cite{DBLP:journals/algorithmica/CherkasskyG97} frequently assigns exact distance labels by performing a reverse BFS from the sink, to reduce relabel work in practice.
Preflows already induce minimum sink-side cuts, so if only cuts are required, the algorithm can already stop once no active nodes with distance label $< n$ exist.

\myparagraph{Flows on Hypergraphs.}
The \emph{Lawler expansion}~\cite{Lawler} of a hypergraph $H = (V,E,c,\omega)$ is a graph consisting of $V$ and two nodes
$\innode, \outnode$ for each $e \in E$, with directed edges $\forall u \in V, e \in \incnets(u): (u, \innode), (\outnode, u)$
with infinite capacity and bridging edges $\forall e \in E: (\innode, \outnode)$ with capacity $\omega(e)$.
A minimum $(\source, \sink)$-cut in the Lawler expansion directly corresponds to one in the hypergraph.

\section{Related Work}\label{sec:related_work}
There is a vast amount of literature on hypergraph partitioning, including extensive surveys~\cite{ALPERT-SURVEY,GRAPH-SURVEY,PAPA-MARKOV,KAHYPAR-DIS}.
The most relevant sequential algorithms are \Partitioner{PaToH}~\cite{PATOH}, \Partitioner{hMetis}~\cite{HMETIS, HMETIS-K}, and \Partitioner{KaHyPar}~\cite{KaHyPar-R, KAHYPAR-K,KAHYPAR-HFC}.
Notable parallel algorithms are \Partitioner{Parkway}~\cite{PARKWAY-2} and \Partitioner{Zoltan}~\cite{ZOLTAN} for distributed memory,
as well as \Partitioner{BiPart}~\cite{BIPART} and \Partitioner{Mt-KaHyPar}~\cite{MT-KAHYPAR,MT-KAHYPAR-Q} for shared memory.

All of these follow the multilevel paradigm that proceeds in three phases:
First, the hypergraph is \emph{coarsened} to obtain a hierarchy of successively smaller and structurally similar
hypergraphs by \emph{contracting} pairs or clusters of vertices.
Once the coarsest hypergraph is small enough, an \emph{initial partition} into $k$ blocks is computed.
Subsequently, the contractions are reverted level-by-level, and, on each level, \emph{local search} heuristics are
used to improve the partition from the previous level (\emph{refinement phase}).



Sanders and Schulz~\cite{KAFFPA} propose an algorithm to improve the edge cut of bipartitions with flow-based refinement.
Their general idea is to grow a size-constrained region around the cut edges of a bipartition.
Afterwards, they compute a minimum $\sourcesink$-cut in the subgraph induced by the region and apply it to the original graph, if it satisfies the balance constraint.
They extended their algorithm to $k$-way partitions by scheduling it on pairs of adjacent blocks.
Heuer~\etal~\cite{KAHYPAR-MF} integrated this approach into their hypergraph partitioner \Partitioner{KaHyPar}.
The framework was then further improved by Gottesbüren~\etal~\cite{KAHYPAR-HFC} by replacing the bipartitioning routine
with the \Partitioner{FlowCutter} algorithm~\cite{REBAHFC,FLOW-CUTTER}.
Flow-based refinement substantially improved the solution quality (connectivity metric) of the partitions produced by
\Partitioner{KaHyPar}, making it the method of choice for high-quality hypergraph partitioning~\cite{KAHYPAR-DIS}.
We will explain the flow-based refinement routine of \Partitioner{KaHyPar} in more detail in the main part of our work.
Notable other flow-based refinement algorithms are \Partitioner{Improve}~\cite{andersen2008algorithm} and \Partitioner{MQI}~\cite{lang2004flow}, which improve the
expansion or conductance metric of bipartitions.


\section{Framework Overview}\label{sec:overview}
\vspace{-0.25cm}
\begin{algorithm2e}[ht]
    \KwIn{Hypergraph $H = (V,E,c,\omega)$, $k$-way partition $\Partition$ of $H$}
    \caption{Parallel Flow-Based Refinement}\label{pseudocode:flow_refinement}

    $\quotientgraph \gets \FuncSty {buildQuotientGraph}(H, \Partition)$ \tcp*[r]{see Section~\ref{sec:scheduling}}  \label{pseudocode:build_quotient_graph}
    \ParallelWhile (\tcp*[f]{see Section~\ref{sec:scheduling}}) {$\exists$ active $(V_i,V_j) \in \quotientgraph$} { \label{pseudocode:active_block_scheduling}
      $B := B_i \cup B_j \gets \FuncSty {constructRegion}(H,V_i,V_j)$ \tcp*[r]{$B_i \subseteq V_i, B_j \subseteq V_j$, see Section~\ref{sec:network_construction}} \label{pseudocode:construct_region}
      $(\flownetwork,\source,\sink) \gets \FuncSty {constructFlowNetwork}(H,B)$ \tcp*[r]{see Section~\ref{sec:network_construction}} \label{pseudocode:construct_flow_network}
      $(M, \Delta_{\text{exp}}) \gets \FuncSty {FlowCutterRefinement}(\flownetwork,\source,\sink)$ \tcp*[r]{see Section~\ref{sec:flowbased_refinement} and~\ref{sec:parallel_flows}} \label{pseudocode:flow_cutter_call}
      \If (\tcp*[f]{potential improvement}) { $\Delta_{\text{exp}} \ge 0$ } {
        $\Delta_{\lambda - 1} \gets \FuncSty {applyMoves}(H,\Partition,M)$ \tcp*[r]{see Section~\ref{sec:scheduling}} \label{pseudocode:apply_moves}
        \lIf (\tcp*[f]{found improvement}) { $\Delta_{\lambda - 1} > 0$ } {
          mark $V_i$ and $V_j$ as active \label{pseudocode:active_block_scheduling_3}
        }
        \lElseIf (\tcp*[f]{no improvement}) {$\Delta_{\lambda - 1} < 0$} { $\FuncSty{revertMoves}(H, \Partition, M)$ \label{pseudocode:revert_moves} }
      }
    }
\end{algorithm2e}

We first give an overview of how the different components in the framework interact with each other, before providing detailed component descriptions in their respective sections.
For this we follow the high level structure of the algorithm, as shown in Algorithm~\ref{pseudocode:flow_refinement}.
We start with a parallel scheduling scheme of block pairs based on the quotient graph in Section~\ref{sec:scheduling}, see line 1 and 2.
For each block pair, we extract a subhypergraph constructed around the boundary nodes of the blocks, which yields a flow network, see line 3 and 4 and Section~\ref{sec:network_construction}.
On each network we run the \Partitioner{FlowCutter} bipartitioning algorithm (line 5), whose partition we convert into a set of moves $M$ and an expected connectivity reduction $\Delta_{\text{exp}}$.
\Partitioner{FlowCutter} and its parallelization are discussed in Section~\ref{sec:flowbased_refinement} and~\ref{sec:parallel_flows} respectively.
If \Partitioner{FlowCutter} claims an improvement, i.e., $\Delta_{\text{exp}} > 0$, we apply the moves to the global partition and compute the exact reduction $\Delta_{\lambda -1}$,
based on which we either mark the blocks for further refinement, or revert the moves, see line~\ref{pseudocode:active_block_scheduling_3} and~\ref{pseudocode:revert_moves}.
We distinguish between expected and actual improvement, due to concurrency conflicts that arise in the scheduler, which is described again in Section~\ref{sec:scheduling}.
Finally, Section~\ref{sec:integration} describes the integration of the flow-based refinement routine into the shared-memory hypergraph partitioner
\Partitioner{Mt-KaHyPar}~\cite{MT-KAHYPAR,MT-KAHYPAR-Q}.
Note that \Partitioner{Mt-KaHyPar} provides data structures to concurrently access and modify the partition $\Partition$ and the connectivity sets $\conset(e)$ and
pin counts $\pinsinpart(e,V_i)$ of each net $e \in E$ and block $V_i \in \Partition$ which we frequently use in the description of the following algorithms.


\section{Parallel Active Block Scheduling}\label{sec:scheduling}

Sanders and Schulz~\cite{KAFFPA} propose the active block scheduling algorithm to apply
their flow-based refinement algorithm for bipartitions on $k$-way partitions.
Their algorithm proceeds in rounds.
In each round, it schedules all pairs of adjacent blocks where at least one is marked as \emph{active}.
Initially, all blocks are marked as active.
If a search on two blocks improves the edge cut, both are marked as active for the next round.

\begin{figure}
  \centering
  \includegraphics[width=0.9\textwidth]{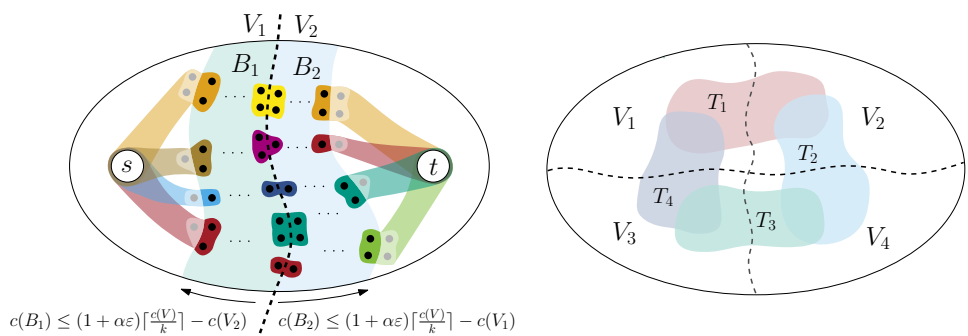}
  \caption{Illustrates the flow network construction algorithm (left) and an how we schedule block pairs of
           the quotient graph $\quotientgraph$ in parallel (right, $T_i$ denotes the search region of thread $i$).}
  \label{fig:flow_network}
  \vspace{-0.5cm}
\end{figure}

\myparagraph{Parallelization.}
A simple parallelization scheme would be to schedule block pairs that form a maximum matching
in the quotient graph $\quotientgraph$ in parallel.
This allows searches to operate in independent
regions of the hypergraph and therefore not to conflict with each other. However, this scheme restricts
the available parallelism to at most $\frac{k}{2}$ threads.
Thus, we do not enforce any constraints on the block pairs processed concurrently, e.g., there can
be multiple threads running on the same block and they can also share some of their nodes
as illustrated in Figure~\ref{fig:flow_network} (right).
This can lead to conflicts when we try to apply a move sequence found by a search on the partition
$\Partition$ and needs to be handled, which we discuss in detail in the next paragraph.

We use $\min(\min(t, \frac{k(k-1)}{2}),\tau \cdot k)$ threads to process the active block pairs in parallel,
where $t$ is the number of available threads in the system and $\frac{k(k-1)}{2}$ is
the maximum number of adjacent blocks in the quotient graph $\quotientgraph$.
The parameter $\tau$ controls the available parallelism in the scheduler.
With higher values of $\tau$, more block pairs are scheduled in parallel.
This can lead to interference between searches that operate on similar regions.
Lower values for $\tau$ can reduce these conflicts but
put more emphasis on good parallelization of $2$-way refinement to achieve good speedups.

Our parallel active block scheduling algorithm uses one global queue $\activequeue$ to schedule active block pairs.
Each active block pair is associated with a round and each round uses an array of size $k$ to mark blocks that become active
in the next round. If a search finds an improvement on two blocks $(V_i,V_j)$
and $V_i$ or $V_j$ becomes active, we push all its
adjacent blocks into $\activequeue$ if they are not already contained in $\activequeue$.
If either $V_i$ or $V_j$ is already active, we insert $(V_i,V_j)$ into $\activequeue$ if not contained.
Thus, active block pairs of different rounds are stored interleaved in $\activequeue$ and the end of each
round does not induce a synchronization point as in the original algorithm~\cite{KAFFPA}.
In the first round, we process all active block pairs in descending order of
improvement they contributed on previous levels.
On ties, we prefer block pairs with a larger number of cut hyperedges.
A round ends when we have processed all its block pairs and all previous rounds have ended.
If the relative improvement at the end of a round is less than $0.1\%$, we immediately terminate the algorithm.

\myparagraph{Apply Moves.}

The flow-based refinement routine returns a move sequence $M$ and the expected reduction
$\Delta_{\text{exp}}$ in the connectivity metric
when we apply $M$ on the partition $\Partition$.
Each move $m \in M$ is of the form $m = (u,V_i,V_j)$ or $m = (u,V_j,V_i)$
which means that node $u$ is moved from block $V_i$ to $V_j$ or vice versa.
If $\Delta_{\text{exp}} \ge 0$, we apply $M$ on the partition $\Partition$.
This may lead to conflicts, since applying $M$ may invalidate some assumptions made by other
threads about the partition $\Partition$ or even the move sequence $M$ itself may be based on some outdated information.

More precisely, there are three conflicts that can occur when we apply $M$.
First, the move sequence leads to a partition that violates the balance constraint.
Second, the actual improvement $\Delta_{\lambda - 1}$ is not equal to the expected
improvement $\Delta_{\text{exp}}$.
This is problematic when we actually degrade the solution quality.
Finally, a node may be in a different block than expected by the move sequence.
The algorithm that applies $M$ is protected via a spin lock
such that no other thread can modify the partition. In practice,
the running time of this algorithm is negligible compared to solving the flow problem and
is therefore not a sequential bottleneck (see Figure~\ref{fig:running_time_shares} in Section~\ref{sec:experiments}).
We first remove all nodes from the sequence that
are not in their expected block. Afterwards, we compute the block weights of the partition
if all remaining moves were applied.
We then perform the moves, if the resulting partition is balanced.
During that, we compute the actual improvement $\Delta_{\lambda - 1}$ of the move sequence.
Suppose we move a node $u$ from block $V_i$ to $V_j$. For each net $e \in I(u)$, we add
$\omega(e)$ to $\Delta_{\lambda - 1}$, if $\pinsinpart(e,V_i)$ decreases to zero and
$-\omega(e)$ to $\Delta_{\lambda - 1}$, if $\pinsinpart(e,V_j)$ increases to one.
If $\Delta_{\lambda - 1} < 0$, we revert all moves.

\myparagraph{Quotient Graph Construction.}
For each block pair, we explicitly store the hyperedges connecting the two. This information is required
by the flow network construction algorithm which we describe in Section~\ref{sec:network_construction}.
Block pairs that contain at least one hyperedge form the edges of the quotient graph.
We construct this data structure by iterating over all hyperedges in parallel
and add a hyperedge $e \in E$ to the block pairs contained in $\{\{V_i,V_j\} \subseteq \conset(e)~\mid~i < j \}$.

If we apply a move sequence on the partition, we add all hyperedges $e \in E$ where $\pinsinpart(e,V_j)$
increases to one to all block pairs contained in $\{\{V_j,V_k\}~\mid~V_k \in \conset(e) \setminus \{V_j\}\}$.
If $\pinsinpart(e,V_i)$ decreases to zero, we remove $e$ lazily from corresponding
block pairs during the flow network construction.

\myparagraph{Implementation Details}
\Partitioner{KaHyPar}~\cite{KAHYPAR-MF} established several pruning rules to skip unpromising flow computations
that we also integrated into our framework. The first rule skips refinement on two adjacent blocks if the cut between
both is less than ten (except on the finest level). The second rule modifies the active block scheduling
algorithm such that after the first round only block pairs are scheduled where at least one computation on them
improved the solution quality on a previous level.

We additionally introduce a time limit to abort long-running flow computations. During scheduling,
we track the average running time $\overbar{t_f}$ required to solve a flow problem and set the time limit
to $8 \cdot \overbar{t_f}$. Note that we activate the time limit
once $k$ (number of blocks) block pairs have been processed.

\section{Network Construction}\label{sec:network_construction}
To improve the cut of a bipartition $\Partition = \{V_1, V_2\}$, we grow a size-constrained region $B$
around the cut hyperedges of $\Partition$. We then contract all nodes in $\overbar{B_1} := V_1 \setminus B$ to the source $\source$ and
$\overbar{B_2} := V_2 \setminus B$ to the sink $\sink$~\cite{KAFFPA,REBAHFC} as illustrated in Figure~\ref{fig:flow_network} (left)
and obtain a coarser hypergraph $\flowhypergraph$.
The flow network $\flownetwork$ is then given by the Lawler expansion of $\flowhypergraph$ (see Section~\ref{sec:preliminaries}).
Note that reducing the hyperedge cut of a bipartition induced by two adjacent blocks of a $k$-way partition $\Partition_k$
optimizes the connectivity metric of $\Partition_k$~\cite{KAHYPAR-MF}.

\myparagraph{Region Growing.}
Sanders and Schulz~\cite{KAFFPA} grow a
region $B := B_1 \cup B_2$ with $B_1 \subseteq V_1$ and $B_2 \subseteq V_2$
around the cut hyperedges of $\Partition$ via two breadth-first-searches (BFS)
as illustrated in Figure~\ref{fig:flow_network} (left).
The first BFS is initialized with all boundary nodes of block $V_1$ and continues to add nodes to $B_1$ as long as
$c(B_1) \le (1 + \alpha \varepsilon) \lceil \frac{c(V)}{2} \rceil - c(V_2)$, where $\alpha$ is an input parameter.
The second BFS that constructs $B_2$ proceeds analogously.
For $\alpha = 1$, each flow computation yields a balanced bipartition
with a possibly smaller cut in the original hypergraph, since only nodes of $B$ can move
to the opposite block ($c(B_1) + c(V_2) \le (1 + \varepsilon)\lceil \frac{c(V)}{2} \rceil$ and vice versa).
Larger values for $\alpha$ lead to larger flow problems with potentially smaller minimum cuts,
but also increase the likelihood of violating the balance constraint.
However, this is not a problem since the flow-based refinement routine guarantees balance through incremental minimum cut computations (see Section~\ref{sec:flowbased_refinement}).


We additionally restrict the distance of each
node $v \in B$ to the cut hyperedges to be smaller than or equal to a parameter $\delta$. We observed that it
is unlikely that a node \emph{far} way from the cut is moved to the opposite block by the flow-based refinement.

\myparagraph{Construction Algorithms.}

We implemented two construction algorithms that are preferable in different situations.
Both construct the hypergraph $\flowhypergraph$ as explained at the beginning of the section.
In the following, we will denote with $E_B := \{e \in E~\mid~e \cap B \neq \emptyset\}$ the set of hyperedges
that contain nodes of $B$.

The first algorithm iterates over all nets $e \in E_B$.
If a pin $p \in e$ is contained in $B$, we add $p$ to hyperedge $e$ in $\flowhypergraph$.
Otherwise, we add the source $\source$ or sink $\sink$ to $e$, if $p \in V_1$ or $p \in V_2$.
The complexity of the algorithm is $\Oh{\sum_{e \in E_B} |e|}$.

The second algorithm iterates over all nodes
$u \in B$ and for each net $e \in I(u)$, we insert a pair $(e,u)$ into a vector.
Sorting the vector (lexicographically) yields the pin lists of the subhypergraph $H_B$.
Afterwards, we insert each net $e$ in the pin list vector into $\flowhypergraph$ and add the source
$\source$ or sink $\sink$ to $e$, if $\pinsinpart(e,B_1) < \pinsinpart(e,V_1)$ or $\pinsinpart(e,B_2) < \pinsinpart(e,V_2)$.
The complexity of the algorithm is $\Oh{p\log{p}}$ where $p := \sum_{e \in E_B} |e \cap B|$ is the number of
pins of $\flowhypergraph$.

The first algorithm has
linear running time, but has to scan all hyperedges of $H$ in their entirety
even if most of their pins are not contained in $\flowhypergraph$.
The complexity of the second algorithm only depends on the number of pins
in $\flowhypergraph$, but requires to sort the pin list vector.
We use the second algorithm for hypergraphs with a low density $d := |E| / |V|$ ($\le 0.5$)
or a large average hyperedge size $\avgsize$ ($\ge 100$).

Note that both algorithms discard single-pin nets and nets that contain both the source and sink
(such nets cannot be removed from the cut).

\myparagraph{Parallelization.}
The first algorithm iterates over all nets $e \in E_B$ in parallel and each thread uses the
sequential algorithm to construct a thread-local pin list vector.
Afterwards, we use a prefix sum operation
to copy the pin lists of each thread to $\flowhypergraph$.

The second algorithm iterates over all nodes $u \in B$ in parallel and then
uses hashing to distribute the pairs $(e,u)$ to buckets.
Afterwards, we process each bucket in parallel and
apply the sequential algorithm to construct the pin list vector of each bucket.
Finally, we use a prefix sum operation to copy the pin lists of each bucket to $\flowhypergraph$.

\myparagraph{Identical Net Removal.}
Since some nets of $H$ are only partially contained in $\flowhypergraph$,
some of them may become identical.
Therefore, we further reduce the size of $\flowhypergraph$ by removing all identical nets
except for one representative at which we aggregate their weight.
We use the identical net detection algorithm of Aykanat~\etal~\cite{INR,kPaToH}.
It uses \emph{fingerprints} $f_e := \sum_{v \in e} v^2$ to eliminate unnecessary pairwise comparisons between nets.
Nets with different fingerprints or different sizes cannot be identical.
If we insert a net $e$ into $\flowhypergraph$, we store the pair $(f_e, e)$ in a hash
table with chaining to resolve collisions
(uses concurrent vectors to handle parallel access).
We can then use the hash table to perform pin-list comparisons
between the nets with the same fingerprint for subsequent net insertions.
Note that in the parallel scenario we may not be able to detect all identical nets due
to simultaneous insertions into the hash table.
However, this does not affect correctness of the refinement, as removing identical nets is only a performance optimization.

\section{Flow-Based Refinement}\label{sec:flowbased_refinement}
In this section we discuss the flow-based refinement routine on a bipartition.
We first introduce the aforementioned \Partitioner{FlowCutter} algorithm~\cite{FLOW-CUTTER, yang-wong-fbb}, which is used as the flow-based refinement routine in \Partitioner{KaHyPar}~\cite{KAHYPAR-HFC}.
The algorithm is parallelized by plugging in a parallel maximum flow algorithm, which we discuss in detail in the next section.
To speed up \Partitioner{FlowCutter}'s convergence and to ensure there is sufficient work to make a parallel algorithm worthwhile, we propose an optimization named \emph{bulk piercing}.

\myparagraph{Core Algorithm.}

\Partitioner{FlowCutter} solves a sequence of incremental maximum flow problems until a balanced bipartition is found.
Algorithm~\ref{pseudocode:flowcutter} shows pseudocode for the approach.
In each iteration, first the previous flow (initially zero) is augmented to a maximum flow regarding the current source set $S$ and sink set $T$, see line 3.
Subsequently, the node sets $S_r, T_r \subset \flownodeset$ of the source- and sink-side cuts are derived in line 4.
This is done via residual BFS (forward from $S$ for $S_r$, backward from $T$ for $T_r$).
The node sets induce two bipartitions $(S_r, \flownodeset \setminus S_r)$ and $(\flownodeset \setminus T_r, T_r)$.
If neither is balanced, all nodes on the side with smaller weight are transformed to a source (if $c(S_r) \leq c(T_r)$) or a sink otherwise, see lines 7-10.
To find a different cut in the next iteration, one additional node is added, called \emph{piercing node}.
Thus, the bipartitions contributed by the currently smaller side will be more balanced in future iterations.
Since the smaller side is grown, this process will converge to a balanced partition.

\begin{algorithm2e}
	\caption{FlowCutter Core}\label{pseudocode:flowcutter}
	$S \gets \{ \source \} , T \gets \{ \sink \}$\;
	\While() {true} {
		augment flow to maximality regarding $S,T$ \;
		derive source- and sink-side cut $S_r, T_r \subset \flownodeset$\;
		\If() {$(S_r, \flownodeset \setminus S_r)$ or $(\flownodeset \setminus T_r, T_r)$ balanced} {
			\Return balanced partition
		}
		\If() {$c(S_r) \leq c(T_r)$} {
			$S \gets S_r \cup \FuncSty{selectPiercingNode()}$
		} \Else() {
			$T \gets T_r \cup \FuncSty{selectPiercingNode()}$
		}
	}
\end{algorithm2e}

\myparagraph{Piercing.}

Selecting a good piercing node is important to achieve good quality, which is why several selection heuristics were proposed.
For our purpose, two heuristics are relevant: \emph{avoiding augmenting paths}~\cite{FLOW-CUTTER, yang-wong-fbb} and \emph{distance from cut}~\cite{KAHYPAR-HFC}.
Whenever possible, a node that is not reachable from the source or sink should be picked, i.e., $v \in \flownodeset \setminus (S_r \cup T_r)$.
Such nodes do not create an augmenting path, and thus the weight of the cut does not increase, while improving balance.
As a secondary criterion, the distance from the original cut is used (larger is better), to prefer reconstructing parts of the original cut over randomized decisions.
This heuristic is only applicable in the refinement setting.


\myparagraph{Most Balanced Cut.}

Once the partition is balanced, we try to improve the balance even further by continuing to pierce as long as no augmenting path is created.
This process is repeated several times with different randomized choices since this is fast (no flow augmentation), starting from the first balanced partition.
More balance gives other refinement algorithms more leeway for improvement.
An equivalent heuristic was already employed in previous flow-based refinement approaches~\cite{KAFFPA, PicardQ82}.

\myparagraph{Bulk Piercing.}

The complexity of \Partitioner{FlowCutter} is $O(\zeta m)$, where $\zeta$ is the final cut weight, and
$m=|\flowedgeset|$.
This bound stems from a pessimistic implementation that augments one flow unit in $O(m)$ work~\cite{FLOW-CUTTER, yang-wong-fbb}.
For refinement, the performance is much better in practice, as the first cut is often close to the final cut.
Only few augmenting iterations are needed and much less than $O(m)$ work is spent per flow unit~\cite{REBAHFC}, with most work spent on the initial flow.

Still, the flow augmented per iteration is often small: at most the capacity of edges incident to the piercing node.
On large instances, we observed that the number of required iterations increases substantially.
We propose to accelerate convergence by piercing multiple nodes per iteration, as long as we cannot avoid augmenting paths and are far from balance.
To ensure a poly-log iteration bound, we set a geometrically shrinking goal of weight to add to each side per iteration.
The initial goal for the source side is set to $\beta (\frac{c(\flownodeset)}{2} - c(\source))$, where $\beta \in (0,1)$ is the geometric shrinking factor that is multiplied with the term in each iteration, and $\frac{c(\flownodeset)}{2} - c(\source)$ is the weight to add for perfect balance.

If a goal is not met, its remainder is added to next iteration's goal.
We track the average weight added per node and from this estimate the number of piercing nodes needed to match the current goal.
To boost measurement accuracy, we pierce one node for the first few rounds.
The sides have distinct measurements and goals, so that we do not pierce too aggressively when the smaller side flips.
This scheme (with $\beta = 0.55$) reduces running time on our largest instances from beyond two hours (time limit) to less than 10 minutes, while not incurring any quality penalties on either small or large instances, as shown in our experiments.

\section{Parallel Maximum Flow Algorithm}\label{sec:parallel_flows}

The maximum flow problem is log-space complete for P~\cite{FlowsPHard}, i.e., the existence of a poly-log depth algorithm is unlikely.
Furthermore, practical algorithms are notoriously difficult to parallelize efficiently~\cite{ShiloachVishkin, BaumstarkSyncPushRelabel, AndersonSetubalPushRelabel, ColoringPushRelabel} and often achieve only mediocre speedups.
Push-relabel algorithms are the most amenable to parallelization~\cite{AndersonSetubalPushRelabel, ColoringPushRelabel, BaumstarkSyncPushRelabel}.
We picked the synchronous parallel algorithm of Baumstark \etal~\cite{BaumstarkSyncPushRelabel} because it restricts the amount of parallel work less than a recent coloring-based algorithm~\cite{ColoringPushRelabel} and does not seem to incur additional work when more threads are added as opposed to recent asynchronous algorithms~\cite{HongHeAsyncPushRelabel}.
The second property is particularly important to us because threads may arbitrarily join a running flow computation due to the work-stealing scheduler.
We first outline their algorithm, then describe a so far undocumented bug followed by our fix, and conclude with implementation details and intricacies of using FlowCutter with preflows.
A preflow already yields a minimum cut, which suffices for our purpose.

\subsection{Synchronous Parallel Push-Relabel}

The algorithm proceeds in rounds in which all active nodes are discharged in parallel.
The flow is updated globally, the nodes are relabeled locally and the excess differences are aggregated in a second array using atomic instructions.
After all nodes have been discharged, the distance labels $\distance$ are updated to the local labels $\distance'$ and the excess deltas are applied.
The discharging operations thus use the labels and excesses from the previous round.
This is repeated until there are no nodes with $\excess(v) > 0$ and $\distance(v) < n$ left.

To avoid concurrently pushing flow on residual arcs in both directions (race condition on flow values), a deterministic winning criterion on the old distance labels is used to determine which direction to push, if both nodes are active.
If a node has an admissible arc it cannot push due to this, it may not be relabeled this round.
Its discharging terminates after finishing the ongoing scan of residual arcs.

The rounds are interleaved with global relabeling~\cite{DBLP:journals/algorithmica/CherkasskyG97}, after linear push and relabel work, using parallel reverse BFS.

\subsection{A Bug in the Synchronous Algorithm}\label{sec:flows:bug}

\begin{figure}
  \centering
  \includegraphics[width=0.9\textwidth]{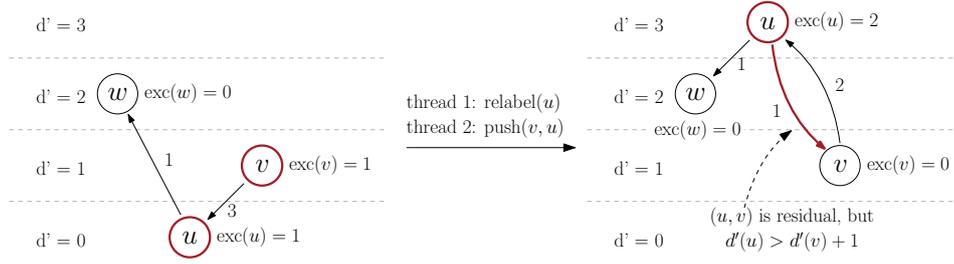}
  \caption{Illustrates a push-relabel conflict in the parallel discharge routine (adapted from Ref.~\cite{ColoringPushRelabel}). The numbers on the arcs denote their residual capacities.}
  \label{fig:push_relabel_bug}
  \vspace{-0.5cm}
\end{figure}

The parallel discharge routine does not protect against push-relabel conflicts~\cite{ColoringPushRelabel} as illustrated in Figure~\ref{fig:push_relabel_bug}.
In particular the winning criterion does not help.
A node $u$ may be relabeled too high if it is concurrently pushed to through a residual arc $(v,u)$ with $\distance'(v) = \distance(u) + 1$.
The arc $(u,v)$ may not be observed as residual yet, and thus $u$ may set its new label $\distance'(u) > \distance'(v) + 1$, violating label correctness.
The bug becomes noticeable when the algorithm terminates prematurely with incorrect distances.

We propose two alternative fixes.
The first is an atomic blocking mechanism, where active nodes are prohibited from being relabeled after being pushed to and prohibited from being pushed to after being relabeled in the same round; whichever operation comes first.
The second fix is to collect mislabeled excess nodes during global relabeling.
When the algorithm would terminate (no active nodes remaining), we run global relabeling, and restart the main loop if new active nodes were found.

We experimentally compare the two methods in~\cref{appendix:flow_algo_comparison}, finding that they perform equivalently well for plain flow computations.
This is because the bug occurs quite rarely, and with the collection of mislabeled excess nodes in intermediate global relabeling steps, the final global relabeling is actually never necessary on our benchmark instances.
For FlowCutter, however, we need the cuts, not just the flow assignment.
Finding the sink-side cut can be done jointly with running the last global relabeling, so its work is already accounted for.
Therefore, we chose this method.

\subsection{Implementation Details}\label{sec:flows:impl-details}

To facilitate an efficient, practical code, we discuss several implementation details.
This covers techniques specific to the hypergraph setting, multi-source multi-sink settings and general techniques.

\myparagraph{Restricting Capacities.}
Recall that only bridge edges $(\innode, \outnode)$ have finite capacity ($\omega(e)$) in the Lawler network.
Since $(\innode, \outnode)$ is the only outgoing edge of $\innode$ with non-zero capacity, the flow (but not preflow) on edges $(u, \innode)$ is also bounded by $\omega(e)$.
Adding these capacities during the preflow stage is a trivial optimization, but it reduces running time for one flow computation on our largest instance from over two hours to 14 seconds, when using 16 cores.
It also boosts the available parallel work, since hypernodes are not immediately relieved of all their excess.
Without this optimization the minimum cut contains only bridge edges, but now may contain edges $(u, \innode)$.
This matters when tracking cut hyperedges (for collecting piercing candidates), which are detected by checking if $\innode$ and $\outnode$ are on different sides.
Therefore, we do not check the capacity and visit $\innode$ nodes during forward residual BFSs.

\myparagraph{Avoid Pushing Flow Back.}
Once the correct flow value is found, the algorithm could terminate in theory.
This is often achieved in very few discharging rounds ($< 1\%$).
Furthermore, we observed that the number of active nodes follows a power law distribution.

At this point flow is only pushed back to the source.
We terminate once all nodes with $\excess(u) > 0$ have $\distance(u) \geq n$, which is most often detected by global relabeling.
Due to little work per round, it takes many rounds to trigger.
We perform additional relabeling, if the flow value has not changed for some rounds (500), and only few active nodes ($< 1500$) were available in each.
Since we expect to terminate, we also set markers for $T_r$.

\myparagraph{Active Nodes.}
The set of active nodes is implemented as an array containing the nodes and an array of insertion timestamps that are atomically set to avoid duplicates.
Nodes are accumulated in thread-local buffers that are frequently flushed to the shared array.
During discharging, we build the array for the next round.
We insert a node if it gets pushed to, or if it has excess left after its discharge operation.
After a round of discharging, we swap the previous active nodes array with the newly built one, and increment the timestamp to reset the markers.

These arrays are reused for global relabeling as well as deriving the source-side and sink-side cuts.
This enables computing the cut-side weights via a simple parallel reduction over the respective arrays, which we use to decide which side to grow.

\myparagraph{Flow Value.}
We track the flow value to abort the refinement in case it exceeds the previous cut.
Traditionally in push-relabel algorithms, the flow value is determined from the excess of the sink.
Since we have many sinks, we do not want to repeatedly accumulate all of their excesses.
Instead, we also insert sinks into the active nodes for the next round.
This way, we can add their excess deltas to the flow value during the update phase, but we of course do not discharge them in the next round.

\myparagraph{Hypergraph Implementation.}
For performance reasons we implement the flow algorithm directly on the hypergraph, simulating the Lawler expansion without actually constructing the graph flow network.
We implement three separate discharge operations that scan the pins plus the bridge edge (in-node and out-node) or the incident hyperedges of a hypernode and push the appropriate amount of flow.
The performance impact of this is demonstrated experimentally in Appendix~\ref{appendix:flow_algo_comparison}.
While it is not as drastic as for Dinitz, where better optimizations are possible~\cite{REBAHFC, KAHYPAR-HFC}, it is still worthwhile.

\subsection{Intricacies with Preflows and FlowCutter}

In this section, we discuss (some unexpected) challenges we faced during the implementation that arose from using FlowCutter with preflows.
The major difference to actual flows is that there are nodes with positive excess left.

\myparagraph{Source-Side Cut.}
First, note that a preflow only yields a sink-side cut via the reverse residual BFS, but for FlowCutter we also need the source-side cut.
We can run a flow decomposition algorithm~\cite{DBLP:journals/algorithmica/CherkasskyG97} to push excess back to the source, to obtain an actual flow and then compute it via forward residual BFS.
However, flow decomposition is difficult to parallelize~\cite{BaumstarkSyncPushRelabel}.
Instead, we initialize the forward residual BFS with all non-sink excess nodes.
This finds the reverse paths that carry flow from the source to the excess nodes, which is precisely what we need.

\myparagraph{Sink-Side Piercing.}
When transforming a node with positive excess to a sink, its excess must be added to the flow value.
Fortunately, this only happens when piercing, as sink-side nodes have no excess if they are not sinks yet.

\myparagraph{Maintain Distance Labels.}
Ideally, we want to reuse the distance labels to avoid re-initialization overheads.
However, as the labels are a lower bound on the distance from the sink, piercing on the sink side invalidates the labels.
Additionally, no new excess nodes are created.
In this case, we run global relabeling to collect the existing excess nodes, before starting the main discharge loop.
When piercing on the source side, new excesses are created, so we do not run the additional global relabel.
Instead, we collect the existing excess nodes during regular runs; at the latest for the termination check.

\section{Integration into a Multilevel Framework}\label{sec:integration}
We integrated our flow-based refinement algorithm in the shared-memory multilevel partitioner \Partitioner{Mt-KaHyPar}.
The framework provides two multilevel partitioners: one opts for
the traditional $\Oh{\log{n}}$ level approach by contracting a vertex clustering on each level (\Partitioner{Mt-KaHyPar-D}~\cite{MT-KAHYPAR})
and one implements a parallel version of the $n$-level scheme~\cite{ADAPTIVE-STOP-RULE,KaHyPar-R,KaHyPar-R}
that (un)contracts only a single vertex on each level (\Partitioner{Mt-KaHyPar-Q}~\cite{MT-KAHYPAR-Q}).

In each refinement step, \Partitioner{Mt-KaHyPar-D} first runs label propagation refinement~\cite{LABEL_PROPAGATION}
followed by FM local search~\cite{FM}.
We run our flow-based refinement as the third component.
\Partitioner{Mt-KaHyPar-Q} uncontracts a fixed number of nodes on each level and then performs a highly-localized
variant of label propagation and FM refinement. Using our flow-based refinement as a third
technique would incur too much overhead, since there are $\Oh{|V|}$ levels and refinement steps. Thus, we use
approximately $\Oh{\log{n}}$ synchronization points similar to \Partitioner{Mt-KaHyPar-D} and perform FM local
search followed by our flow-based refinement.
We do not run label propagation here since there were no quality benefits.
We run all refinement algorithms on each level multiple times in combination and stop if the relative improvement is less than $0.25\%$.

\section{Experiments}\label{sec:experiments}
We implemented the flow-based refinement routine in the shared-memory hypergraph partitioner \Partitioner{Mt-KaHyPar}\footnote{\mtkahypar~is available from \url{https://github.com/kahypar/mt-kahypar}},
which is implemented in \texttt{C++17}, parallelized using the TBB library~\cite{TBB}, and compiled using \gpp{9.2} with the
flags \texttt{-O3 -mtune=native -march=native}. \Partitioner{Mt-KaHyPar} provides two partitioners: \Partitioner{Mt-KaHyPar-D} (\textbf{D}efault setting)
and \Partitioner{Mt-KaHyPar-Q} (\textbf{Q}uality setting). We refer to the corresponding versions that use flow-based refinement as
\mtkahypardflows~and \mtkahyparqflows~(-\textbf{F}lows).
For parallel partitioners we add a suffix to their name to indicate the number of threads used, e.g. \mtkahyparqflowsconfig{64} for 64 threads.
We omit the suffix for sequential partitioners.

\myparagraph{Instances.}

\begin{figure}[!t]
	\centering
  \ifpdfplots
    \includegraphics{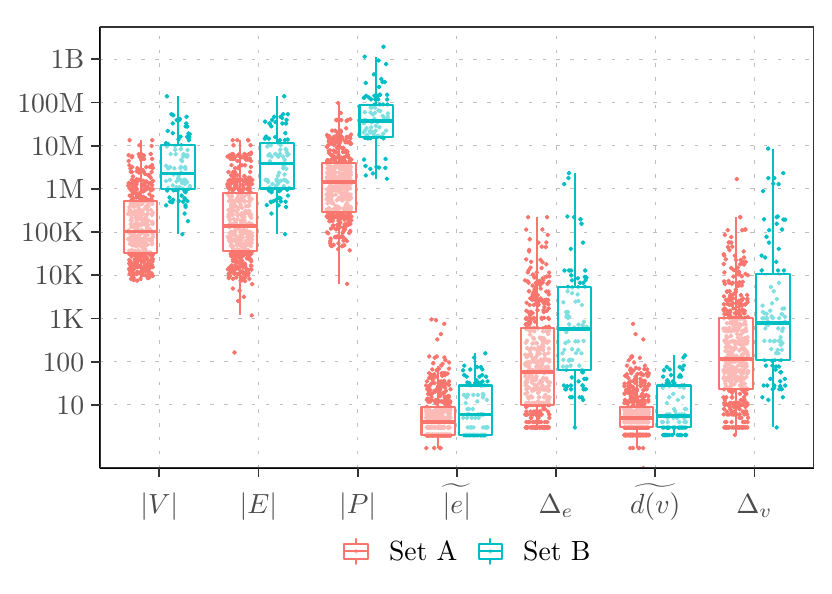}
  \fi
	\vspace{-0.4cm}
  \caption{Summary of different properties for our two benchmark sets. It shows for each
           hypergraph (points), the number of vertices $|V|$, nets $|E|$ and pins $|P|$, as well as the median and maximum
           net size ($\medsize$ and $\maxsize{e}$) and vertex degree ($\meddeg$ and $\maxsize{v}$).}
	\label{fig:benchmark_set}
\end{figure}

All instances of the benchmark sets used in the experimental evaluation are derived from four sources encompassing three application domains:
the ISPD98 VLSI Circuit Benchmark Suite~\cite{ISPD98}, the DAC 2012 Routability-Driven Placement Contest~\cite{DAC}, the SuiteSparse Matrix Collection~\cite{SPM},
and the 2014 SAT Competition~\cite{SAT14}. VLSI instances are transformed into hypergraphs by converting the netlist of each circuit into a set of hyperedges.
Sparse matrices are translated to hypergraphs using the row-net model~\cite{PATOH} and SAT instances to three different hypergraph representations:
\emph{literal}, \emph{primal}, and \emph{dual}~\cite{MANN-PAPA14, PAPA-MARKOV} (see~\cite{KAHYPAR-CA} for more details).
All hypergraphs have unit vertex and net weights.

For comparison with sequential partitioners, we use the established benchmark set of Heuer and Schlag~\cite{KAHYPAR-CA} (referred to as set A, $488$ hypergraphs).
To measure speedups and to compare our implementation with other parallel partitioners,
we use a benchmark set composed of $94$ large hypergraphs (referred to as set B) that was initially assembled to evaluate
\Partitioner{\mtkahyparold}~\cite{MT-KAHYPAR}.
Figure~\ref{fig:benchmark_set} shows that the hypergraphs of set B are more than an order of magnitude larger than those of
set A\footnote{The benchmark sets and experimental results are available from \url{https://algo2.iti.kit.edu/heuer/sea22}}.

\myparagraph{Setup.}
On set A, we use $k \in \{2,4,8,16,32,64,128\}$, $\varepsilon = 0.03$, ten different seeds and a time limit of eight hours.
The experiments are done on a cluster of Intel Xeon Gold 6230 processors ($2$ sockets with $20$ cores each) running at $2.1$ GHz with $96$GB RAM (machine A).

On set B, we use $k \in \{2,8,16,64\}, \varepsilon = 0.03$, three seeds, and a time limit of two hours.
These experiments are run on an AMD EPYC Rome 7702P ($1$ socket with $64$ cores) running at $2.0$--$3.35$ GHz with $1024$GB RAM (machine B).
The parameter space on set B is restricted, since we only have access to one machine of type B.


\myparagraph{Methodology.}

Each partitioner optimizes the connectivity metric, which we also refer to as the quality of a partition.
For each instance (hypergraph and $k$), we aggregate running times using the arithmetic mean over all seeds.
To further aggregate over multiple instances, we use the geometric mean for absolute running times and self-relative speedups.
For runs that exceeded the time limit, we use the time limit itself in the aggregates.
In plots, we mark these instances with \ClockLogo~if \emph{all} runs of that algorithm timed out.

To compare the solution quality of different algorithms, we use \emph{performance profiles}~\cite{PERFORMANCE-PROFILES}.
Let $\mathcal{A}$ be the set of algorithms we want to compare, $\mathcal{I}$ the set of instances, and $q_{A}(I)$ the quality of algorithm
$A \in \mathcal{A}$ on instance $I \in \mathcal{I}$.
For each algorithm $A$, we plot the fraction of instances ($y$-axis) for which $q_A(I) \leq \tau \cdot \min_{A' \in \mathcal{A}}q_{A'}(I)$, where $\tau$ is on the $x$-axis.
Achieving higher fractions at lower $\tau$-values is considered better.
For $\tau = 1$, the $y$-value indicates the percentage of instances for which an algorithm performs best.


\myparagraph{Parameter Configuration.}
We performed extensive parameter tuning experiments which we summarize in detail in Appendix~\ref{appendix:parameter_tuning}.
The scheduler uses $\min(t,k)$ threads ($\tau = 1$) to process adjacent blocks of the quotient graph in parallel.
We also enable bulk piercing as it has no impact on the solution quality while being consistently faster, with the geometric shrinking factor $\beta$ set to $0.55$.
Further, we restrict the distance of each node to the cut hyperedges when we grow the region $B$ to be smaller than or equal to $\delta = 2$.
Finally, we set the region scaling parameter $\alpha = 16$ which is also used in
the flow-based refinement algorithm of \Partitioner{KaHyPar}~\cite{KAHYPAR-MF}.

\subsection{Comparison with other Algorithms}

We now compare different partitioners with \Partitioner{Mt-KaHyPar} when using flow-based refinement.
Since performance profiles do not permit a full ranking of three or more algorithms,
we additionally add pairwise comparisons of \mtkahyparqflows~and all evaluated partitioners using performance profiles
in Appendix~\ref{appendix:pairwise_comparisons}.

\begin{figure*}[t]
  \begin{minipage}{.99\textwidth}
  \ifpdfplots
    \includegraphics{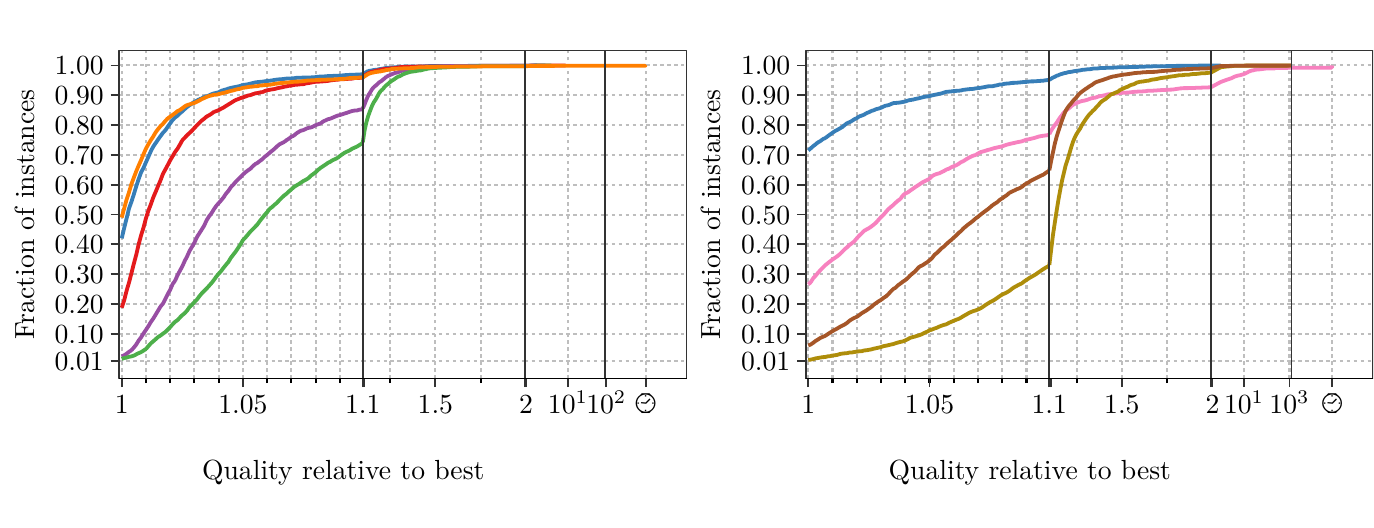}
  \fi
  \end{minipage} %
  \begin{minipage}{.99\textwidth}
    \vspace{-0.65cm}
    \centering
  \ifpdfplots
    \includegraphics{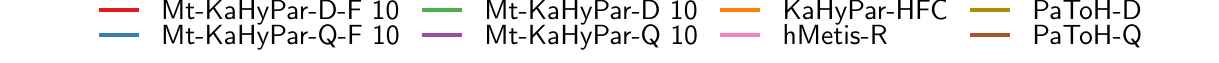}
  \fi
  \end{minipage} %
	\vspace{-0.25cm}
  \caption{Performance profiles comparing solution quality of \mtkahyparqflows~with different partitioners on set A.}
  \label{fig:quality_set_a}
\end{figure*}

\myparagraph{Medium-Sized Instances.}

\begin{filecontents*}{data/set_a.dat}
betterThanPaToHD = 94.7
betterThanPaToHQ = 87.7
betterThanMtKaHyParD = 97.3
betterThanHMetisR = 74.5
betterThanMtKaHyParQ = 95.7
betterThanMtKaHyParHD = 73.5
betterThanKaHyParHFC = 51.3
medianImprovementDefaultFlowsVsNonFlows = 4.2
medianImprovementQualityFlowsVsNonFlows = 2.7
medianImprovementQualityFlowsVsDefaultFlows = 0.6
medianImprovementPaToHQVsPaToHD = 5.3
medianImprovementHMetisRVsPaToHQ = 2.6
medianImprovementQualityFlowsVsHMetisR = 3
medianImprovementMtKaHyParQVsD = 1.9
runtimeFactorDefaultFlowsVsNonFlows = 3.1
runtimeFactorQualityFlowsVsNonFlows = 1.7
runtimeFactorQualityFlowsVsDefaultFlows = 1.9
runtimeFactorPaToHQVsPaToHD = 5
runtimeFactorHMetisRVsPaToHQ = 15.9
runtimeFactorQualityFlowsVsHMetisR = 18.3
runtimeFactorQualityFlowsVsKaHyParHFC = 9.6
gmean_mt_kahypar_hq = 5.08
gmean_mt_kahypar_hd = 2.73
gmean_mt_kahypar_q = 2.99
gmean_mt_kahypar_d = 0.89
gmean_patoh_d = 1.17
gmean_patoh_q = 5.86
gmean_hmetis_r = 93.21
gmean_kahypar_hfc = 48.98
\end{filecontents*}
\DTLloaddb[noheader, keys={key,value}]{set_a}{data/set_a.dat}

On set A, we compare \Partitioner{Mt-KaHyPar} to \Partitioner{KaHyPar-HFC}~\cite{KAHYPAR-MF, REBAHFC}
(uses similar algorithmic components as \mtkahyparqflows) which is the current best sequential
partitioner in terms of solution quality~\cite{KAHYPAR-DIS}, the recursive bipartitioning version (\Partitioner{hMetis-R}) of
\Partitioner{hMetis $2.0$}~\cite{HMETIS}, as well as the default (\Partitioner{PaToH-D}) and quality preset (\Partitioner{PaToH-Q}) of
\Partitioner{PaToH $3.3$}~\cite{PATOH}. All configurations of \Partitioner{Mt-KaHyPar} use $10$ threads.

\begin{figure*}[t]
  \centering
  \ifpdfplots
    \includegraphics{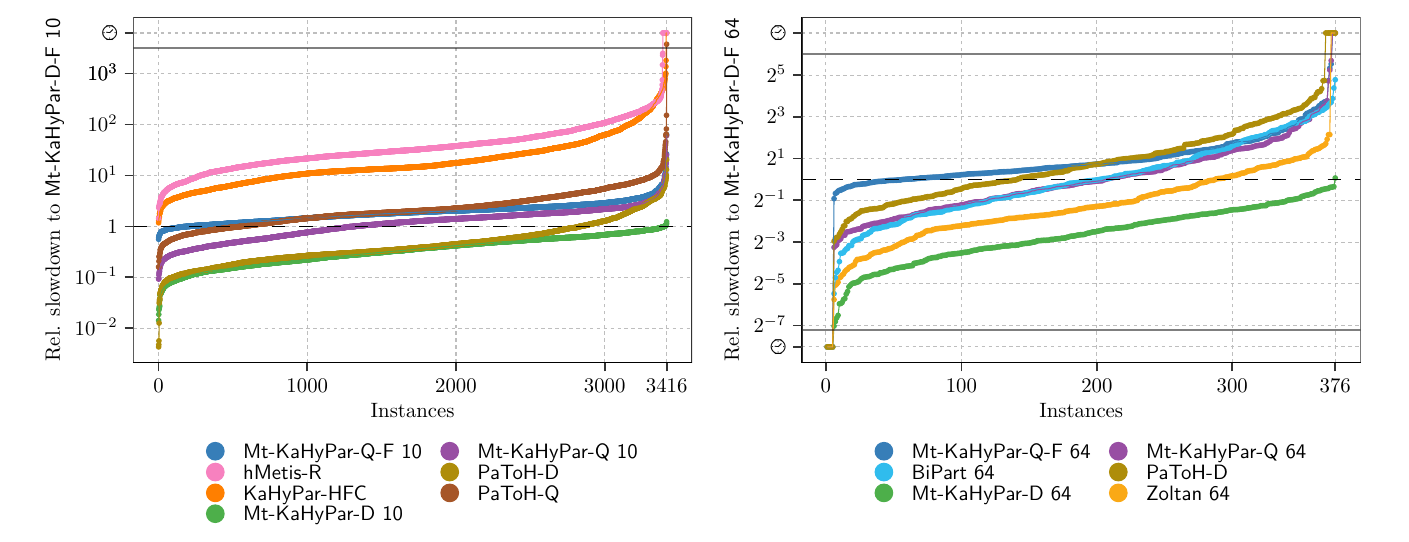}
  \fi
	\vspace{-0.6cm}
  \caption{Running times relative to \mtkahypardflows~on set A (left) and B (right).
           The \ClockLogo~axis markers represent timouts for the baseline \mtkahypardflows~(at the bottom)
           or the compared algorithm (at the top).}
  \label{fig:relative_running_time}
\end{figure*}

Figure~\ref{fig:quality_set_a} and~\ref{fig:relative_running_time} (left) compare the solution quality and running times of \Partitioner{Mt-KaHyPar} with different partitioners on set A.
In an individual comparison, \mtkahyparqflows~finds better partitions than \Partitioner{PaToH-D}, \Partitioner{PaToH-Q},
\Partitioner{Mt-KaHyPar-D}, \Partitioner{Mt-KaHyPar-Q}, \mtkahypardflows, \Partitioner{hMetis-R} and \Partitioner{KaHyPar-HFC}
on $\placeholder{set_a}{betterThanPaToHD}\%$, $\placeholder{set_a}{betterThanPaToHQ}\%$, $\placeholder{set_a}{betterThanMtKaHyParD}\%$,
$\placeholder{set_a}{betterThanMtKaHyParQ}\%$, $\placeholder{set_a}{betterThanMtKaHyParHD}\%$, $\placeholder{set_a}{betterThanHMetisR}\%$
and $\placeholder{set_a}{betterThanKaHyParHFC}\%$ of the instances, respectively.

The median improvement of \mtkahypardflows~and \mtkahyparqflows~compared to the configurations that use
no flow-based refinement is $\placeholder{set_a}{medianImprovementDefaultFlowsVsNonFlows}\%$ and
$\placeholder{set_a}{medianImprovementQualityFlowsVsNonFlows}\%$ while only incuring a slowdown by
a factor of $\placeholder{set_a}{runtimeFactorDefaultFlowsVsNonFlows}$
(gmean time $\placeholder{set_a}{gmean_mt_kahypar_hd}$s vs $\placeholder{set_a}{gmean_mt_kahypar_d}$s) and
$\placeholder{set_a}{runtimeFactorQualityFlowsVsNonFlows}$
($\placeholder{set_a}{gmean_mt_kahypar_hq}$s vs $\placeholder{set_a}{gmean_mt_kahypar_q}$s).
To put this into perspective, the quality preset of \Partitioner{PaToH} (\Partitioner{PaToH-Q})
improves the default preset (\Partitioner{PaToH-D}) by
$\placeholder{set_a}{medianImprovementPaToHQVsPaToHD}\%$ in the median and is a factor
of $\placeholder{set_a}{runtimeFactorPaToHQVsPaToHD}$ slower
($\placeholder{set_a}{gmean_patoh_q}$s vs $\placeholder{set_a}{gmean_patoh_d}$s).
The median improvement of \Partitioner{hMetis-R} compared to \Partitioner{PaToH-Q}
is $\placeholder{set_a}{medianImprovementHMetisRVsPaToHQ}\%$ while it is a factor
of $\placeholder{set_a}{runtimeFactorHMetisRVsPaToHQ}$ slower
($\placeholder{set_a}{gmean_hmetis_r}$s vs $\placeholder{set_a}{gmean_patoh_q}$s).
The solutions produced by \mtkahyparqflows~are $\placeholder{set_a}{medianImprovementQualityFlowsVsHMetisR}\%$
better than those of \Partitioner{hMetis-R} in the median and it has a similar running time as \Partitioner{PaToH-Q}
($\placeholder{set_a}{gmean_mt_kahypar_hq}$s vs $\placeholder{set_a}{gmean_patoh_q}$s).
If we compare our two partitioners that use flow-based refinement (see also Figure~\ref{fig:quality_pairwise_set_a} in Appendix~\ref{appendix:pairwise_comparisons}), we can see that
\mtkahyparqflows~gives only minor quality improvements over \mtkahypardflows~(median improvement
is $\placeholder{set_a}{medianImprovementQualityFlowsVsDefaultFlows}\%$ whereas without flow-based refinement
it is $\placeholder{set_a}{medianImprovementMtKaHyParQVsD}\%$).
This demonstrates the effectiveness of flow-based refinement.
The solution quality of \mtkahyparqflows~and \Partitioner{KaHyPar-HFC} are on par (see also Figure~\ref{fig:quality_pairwise_set_a} in Appendix~\ref{appendix:pairwise_comparisons}),
while \mtkahyparqflows~is an order of magnitude faster with $10$ threads
($\placeholder{set_a}{gmean_mt_kahypar_hq}$s vs $\placeholder{set_a}{gmean_kahypar_hfc}$s).
In conclusion, we achieved the solution quality of the currently hiqhest quality sequential partitioner in a fast parallel code.

\begin{figure*}[t]
  \vspace{-0.65cm}
  \begin{minipage}{.99\textwidth}
  \ifpdfplots
    \includegraphics{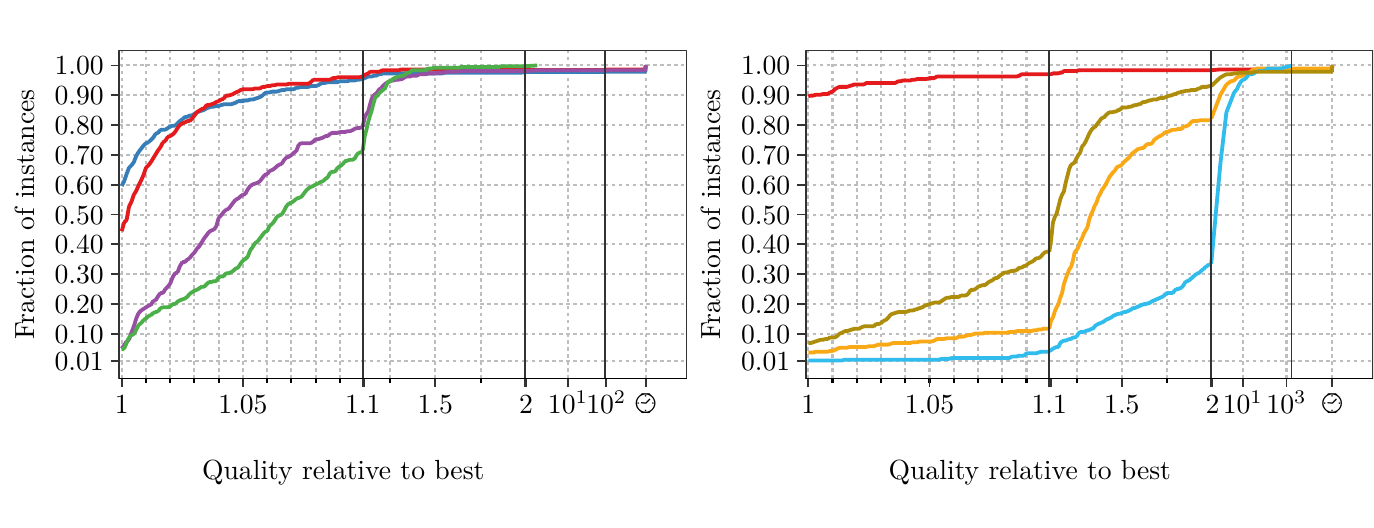}
  \fi
  \end{minipage} %
  \begin{minipage}{.99\textwidth}
    \vspace{-0.65cm}
    \centering
  \ifpdfplots
    \includegraphics{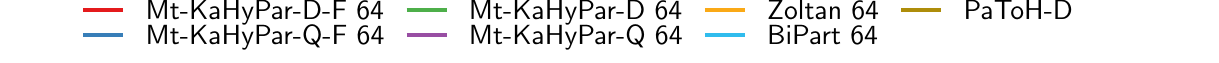}
  \fi
  \end{minipage} %
	\vspace{-0.25cm}
  \caption{Performance profiles comparing solution quality of \mtkahypardflows~with different partitioners on set B.}
  \label{fig:quality_set_b}
\end{figure*}

\myparagraph{Large Instances.}

\begin{filecontents*}{data/set_b.dat}
betterThanPaToHD = 91.8
betterThanZoltan = 94.7
betterThanBipart = 96.5
betterThanMtKaHyParD = 89.1
betterThanMtKaHyParQ = 88.8
betterThanMtKaHyParHD = 62.8
defaultBetterThanPaToHD = 92
defaultBetterThanZoltan = 96
defaultBetterThanBipart = 97.3
defaultBetterThanMtKaHyParD = 91
defaultBetterThanMtKaHyParQ = 85.1
defaultBetterThanMtKaHyParHQ = 47.3
medianImprovementDefaultFlowsVsNonFlows = 5.2
medianImprovementQualityFlowsVsNonFlows = 3.4
medianImprovementQualityFlowsVsZoltan = 34
medianImprovementQualityFlowsVsPaToHD = 14
medianImprovementQualityFlowsVsBipart = 200
medianImprovementDefaultFlowsVsZoltan = 33
medianImprovementDefaultFlowsVsPaToHD = 13
medianImprovementDefaultFlowsVsBipart = 200
runtimeFactorDefaultFlowsVsNonFlows = 6.6
runtimeFactorQualityFlowsVsNonFlows = 1.9
runtimeFactorQualityFlowsVsDefaultFlows = 1.9
runtimeFactorDefaultFlowsVsZoltan = 2.4
runtimeFactorDefaultFlowsVsBipart = 1
runtimeFactorDefaultFlowsVsPaToHD = 0.6
gmean_mt_kahypar_hq = 58.24
gmean_mt_kahypar_hd = 30.38
gmean_mt_kahypar_q = 29.99
gmean_mt_kahypar_d = 4.63
gmean_patoh_d = 50.3
gmean_zoltan = 12.6
gmean_bipart = 29.19
\end{filecontents*}
\DTLloaddb[noheader, keys={key,value}]{set_b}{data/set_b.dat}

On set B, we compare \Partitioner{Mt-KaHyPar} with the parallel algorithms \Partitioner{Zoltan} $3.83$~\cite{ZOLTAN} and \Partitioner{BiPart}~\cite{BIPART},
as well as \Partitioner{PaToH-D}, which is the only sequential algorithm to complete the experiments in a reasonable time frame.
All parallel algorithms use $64$ threads.

Figure~\ref{fig:quality_set_b} and~\ref{fig:relative_running_time} (right) compare the solution quality and running times of \Partitioner{Mt-KaHyPar} with different partitioners on set B.
The quality of the partitons produced by \mtkahypardflows~and \mtkahyparqflows~are comparable
(see also Figure~\ref{fig:quality_pairwise_set_b} in Appendix~\ref{appendix:pairwise_comparisons}) while
\mtkahypardflows~is a factor of $\placeholder{set_b}{runtimeFactorQualityFlowsVsDefaultFlows}$ faster
(gmean time $\placeholder{set_b}{gmean_mt_kahypar_hd}$s vs $\placeholder{set_b}{gmean_mt_kahypar_hq}$s).
Therefore, we focus on \mtkahypardflows~in this evaluation.
In an individual comparison, \mtkahypardflows~finds better partitions than \Partitioner{BiPart}, \Partitioner{Zoltan},
\Partitioner{PaToH-D}, \Partitioner{Mt-KaHyPar-D}, \Partitioner{Mt-KaHyPar-Q} and \mtkahyparqflows~on
$\placeholder{set_b}{defaultBetterThanBipart}\%$, $\placeholder{set_b}{defaultBetterThanZoltan}\%$, $\placeholder{set_b}{defaultBetterThanPaToHD}\%$,
$\placeholder{set_b}{defaultBetterThanMtKaHyParD}\%$, $\placeholder{set_b}{defaultBetterThanMtKaHyParQ}\%$ and $\placeholder{set_b}{defaultBetterThanMtKaHyParHQ}\%$
of the instances, respectively.

The median improvement of \mtkahypardflows~and \mtkahyparqflows~compared to the configurations that use
no flow-based refinement is $\placeholder{set_b}{medianImprovementDefaultFlowsVsNonFlows}\%$ and
$\placeholder{set_b}{medianImprovementQualityFlowsVsNonFlows}\%$ while they are slower by a factor
of $\placeholder{set_b}{runtimeFactorDefaultFlowsVsNonFlows}$
($\placeholder{set_b}{gmean_mt_kahypar_hd}$s vs $\placeholder{set_b}{gmean_mt_kahypar_d}$s) and
$\placeholder{set_b}{runtimeFactorQualityFlowsVsNonFlows}$
($\placeholder{set_b}{gmean_mt_kahypar_hq}$s vs $\placeholder{set_b}{gmean_mt_kahypar_q}$s).
Both the improvements and slowdowns are more pronounced here than on set A.
The slowdowns are expected since the size of the flow problems scales linearly with instance sizes, while the
complexity of the flow-based refinement routine does not.
\mtkahypardflows~($\placeholder{set_b}{gmean_mt_kahypar_hd}$s) is slower
than \Partitioner{Zoltan} ($\placeholder{set_b}{gmean_zoltan}$s) and
\Partitioner{BiPart} ($\placeholder{set_b}{gmean_bipart}$s), but faster
than \Partitioner{PaToH-D} ($\placeholder{set_b}{gmean_patoh_d}$s).
However, \mtkahypardflows~computes partitions that are
$\placeholder{set_b}{medianImprovementDefaultFlowsVsZoltan}\%$
better than those of \Partitioner{Zoltan} and twice as good as
those of \Partitioner{BiPart} in the median.

\subsection{Scalability}\label{sec:experiments:scalability}

\begin{filecontents*}{data/scalability.dat}
gmean_mt_kahypar_hd_1 = 163.96
gmean_mt_kahypar_hd_4 = 52.88
gmean_mt_kahypar_hd_16 = 22.17
gmean_mt_kahypar_hd_64 = 15.44
max_speedup_mt_kahypar_4 = 33.37
max_speedup_mt_kahypar_16 = 41.52
max_speedup_mt_kahypar_64 = 77.21
gmean_speedup_mt_kahypar_4 = 3.1
gmean_speedup_mt_kahypar_16 = 7.4
gmean_speedup_mt_kahypar_64 = 10.62
gmean_speedup_large_mt_kahypar_4 = 3.12
gmean_speedup_large_mt_kahypar_16 = 8.38
gmean_speedup_large_mt_kahypar_64 = 14.5
gmean_speedup_mt_kahypar_4_k2 = 3.45
gmean_speedup_mt_kahypar_16_k2 = 6.37
gmean_speedup_mt_kahypar_64_k2 = 7.56
gmean_speedup_large_mt_kahypar_4_k2 = 4.43
gmean_speedup_large_mt_kahypar_16_k2 = 8.76
gmean_speedup_large_mt_kahypar_64_k2 = 12.05
gmean_speedup_mt_kahypar_4_k8 = 2.54
gmean_speedup_mt_kahypar_16_k8 = 6.25
gmean_speedup_mt_kahypar_64_k8 = 8.68
gmean_speedup_large_mt_kahypar_4_k8 = 2.37
gmean_speedup_large_mt_kahypar_16_k8 = 7.59
gmean_speedup_large_mt_kahypar_64_k8 = 11.86
gmean_speedup_mt_kahypar_4_k16 = 3
gmean_speedup_mt_kahypar_16_k16 = 6.6
gmean_speedup_mt_kahypar_64_k16 = 9.47
gmean_speedup_large_mt_kahypar_4_k16 = 2.85
gmean_speedup_large_mt_kahypar_16_k16 = 5.93
gmean_speedup_large_mt_kahypar_64_k16 = 11.93
gmean_speedup_mt_kahypar_4_k64 = 3.52
gmean_speedup_mt_kahypar_16_k64 = 11.38
gmean_speedup_mt_kahypar_64_k64 = 20.48
gmean_speedup_large_mt_kahypar_4_k64 = 3.45
gmean_speedup_large_mt_kahypar_16_k64 = 11.17
gmean_speedup_large_mt_kahypar_64_k64 = 20.59
max_speedup_mt_kahypar_flow_4 = 63.8
max_speedup_mt_kahypar_flow_16 = 67.78
max_speedup_mt_kahypar_flow_64 = 80.05
gmean_speedup_mt_kahypar_flow_4 = 2.73
gmean_speedup_mt_kahypar_flow_16 = 5.28
gmean_speedup_mt_kahypar_flow_64 = 6.75
gmean_speedup_large_mt_kahypar_flow_4 = 3.04
gmean_speedup_large_mt_kahypar_flow_16 = 7.77
gmean_speedup_large_mt_kahypar_flow_64 = 13.27
gmean_speedup_mt_kahypar_flow_4_k2 = 2.65
gmean_speedup_mt_kahypar_flow_16_k2 = 3.03
gmean_speedup_mt_kahypar_flow_64_k2 = 2.94
gmean_speedup_large_mt_kahypar_flow_4_k2 = 5.98
gmean_speedup_large_mt_kahypar_flow_16_k2 = 9.93
gmean_speedup_large_mt_kahypar_flow_64_k2 = 13.32
gmean_speedup_mt_kahypar_flow_4_k8 = 2.23
gmean_speedup_mt_kahypar_flow_16_k8 = 4.41
gmean_speedup_mt_kahypar_flow_64_k8 = 5.75
gmean_speedup_large_mt_kahypar_flow_4_k8 = 2.21
gmean_speedup_large_mt_kahypar_flow_16_k8 = 7.64
gmean_speedup_large_mt_kahypar_flow_64_k8 = 11.74
gmean_speedup_mt_kahypar_flow_4_k16 = 2.77
gmean_speedup_mt_kahypar_flow_16_k16 = 5.45
gmean_speedup_mt_kahypar_flow_64_k16 = 6.67
gmean_speedup_large_mt_kahypar_flow_4_k16 = 2.7
gmean_speedup_large_mt_kahypar_flow_16_k16 = 5.31
gmean_speedup_large_mt_kahypar_flow_64_k16 = 11.05
gmean_speedup_mt_kahypar_flow_4_k64 = 3.39
gmean_speedup_mt_kahypar_flow_16_k64 = 10.66
gmean_speedup_mt_kahypar_flow_64_k64 = 18.48
gmean_speedup_large_mt_kahypar_flow_4_k64 = 3.24
gmean_speedup_large_mt_kahypar_flow_16_k64 = 10.05
gmean_speedup_large_mt_kahypar_flow_64_k64 = 17.61
\end{filecontents*}
\DTLloaddb[noheader, keys={key,value}]{scalability}{data/scalability.dat}
\begin{filecontents*}{data/conflicts.dat}
med_found_improvement_t1 = 31.83
med_found_improvement_t4 = 31.91
med_found_improvement_t16 = 33.18
med_found_improvement_t64 = 33.38
med_success_t1 = 100
med_success_t4 = 96.73
med_success_t16 = 90.48
med_success_t64 = 85.62
med_balance_violations_t1 = 0
med_balance_violations_t4 = 0.77
med_balance_violations_t16 = 2
med_balance_violations_t64 = 2.55
med_degrade_solution_t1 = 0
med_degrade_solution_t4 = 1.72
med_degrade_solution_t16 = 4.74
med_degrade_solution_t64 = 8.06
\end{filecontents*}
\DTLloaddb[noheader, keys={key,value}]{conflicts}{data/conflicts.dat}
\begin{filecontents*}{data/flow_cutter.dat}
max_speedup_flow_cutter_4 = 4.1
max_speedup_flow_cutter_16 = 10.4
max_speedup_flow_cutter_64 = 18.4
gmean_speedup_flow_cutter_4 = 3
gmean_speedup_flow_cutter_16 = 4.2
gmean_speedup_flow_cutter_64 = 4.4
gmean_speedup_flow_cutter_long_4 = 3.1
gmean_speedup_flow_cutter_long_16 = 6.5
gmean_speedup_flow_cutter_long_64 = 8.6
\end{filecontents*}
\DTLloaddb[noheader, keys={key,value}]{flow_cutter}{data/flow_cutter.dat}
\begin{filecontents*}{data/plain_flows.dat}
max_speedup_plain_flows_4 = 4
max_speedup_plain_flows_16 = 10.3
max_speedup_plain_flows_64 = 17.3
gmean_speedup_plain_flows_4 = 3.1
gmean_speedup_plain_flows_16 = 4.4
gmean_speedup_plain_flows_64 = 4.7
gmean_speedup_plain_flows_long_4 = 3.2
gmean_speedup_plain_flows_long_16 = 7
gmean_speedup_plain_flows_long_64 = 9.9
\end{filecontents*}
\DTLloaddb[noheader, keys={key,value}]{plain_flows}{data/plain_flows.dat}

\begin{figure*}[!t]
  \begin{minipage}{.99\textwidth}
  \ifpdfplots
    \includegraphics{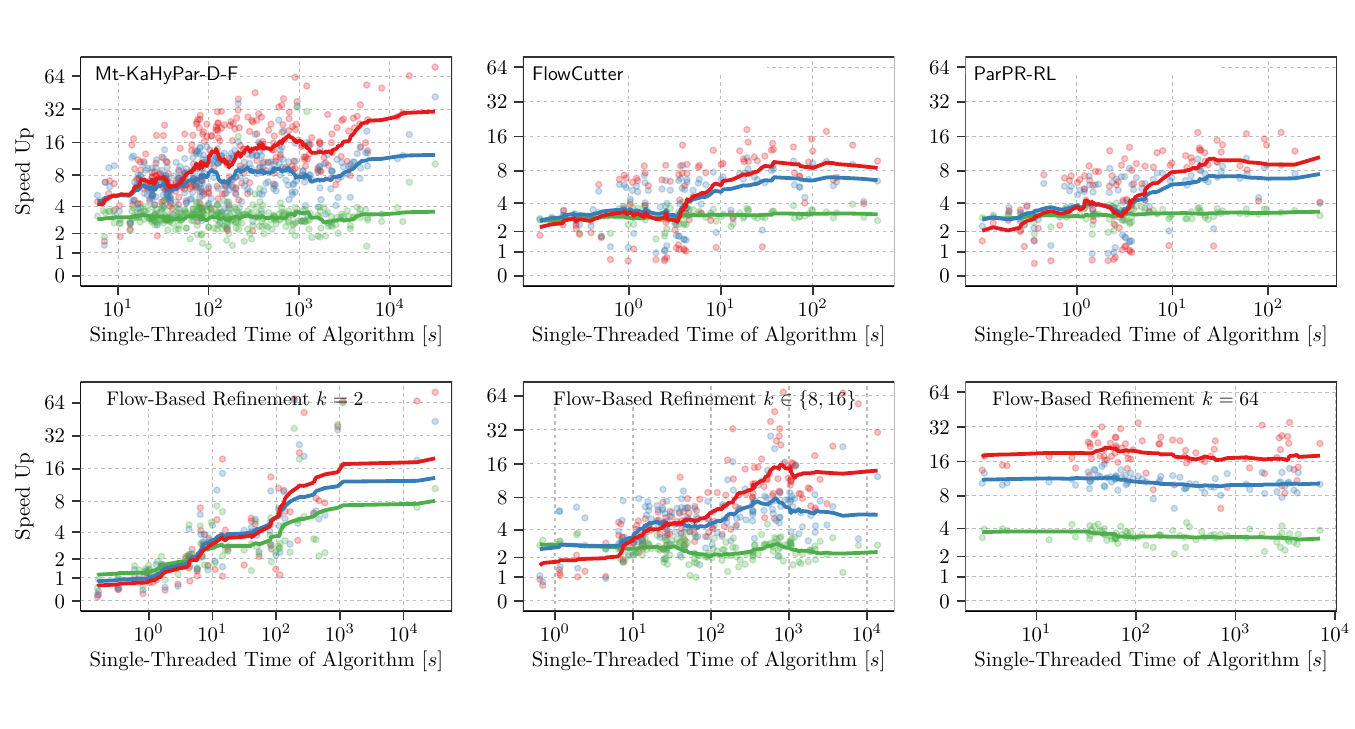}
  \fi
  \end{minipage} %
  \begin{minipage}{.99\textwidth}
    \vspace{-0.65cm}
    \centering
  \ifpdfplots
    \includegraphics{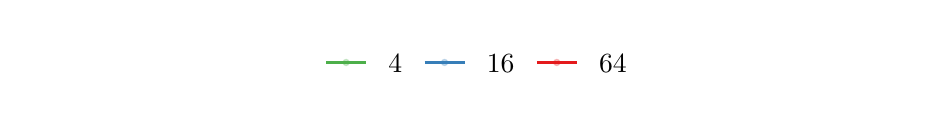}
  \fi
  \end{minipage} %
	\vspace{-0.5cm}
  \caption{Speedups of \mtkahypardflows~and the flow-based refinement routine (for different values of $k$) as well as of the
           \Partitioner{FlowCutter} and parallel flow algorithm (\Partitioner{ParPR-RL}).}
  \label{fig:speedups}
\end{figure*}

In Figure~\ref{fig:speedups}, we summarize self-relative speedups for several algorithmic
components of \mtkahypardflows~with varying number of threads $t \in \{4,16,64\}$.
In the plot, we represent the speedup of each instance as a point and the centered rolling
geometric mean with a window size of $25$ as a line.

\myparagraph{FlowCutter.}

To assess the scalability of \Partitioner{FlowCutter} and the flow algorithm \Partitioner{ParPR-RL}, we extract flow networks from bipartitions of the instances in set B.
The instances are available on the website along the other benchmark instances.
The results are shown in the top middle and right plots of Figure~\ref{fig:speedups}.
With 4 threads, we observe near-perfect speedups throughout, with fairly small variance.
For $t = 16,64$, the parallelization overheads are only outweighed for longer running instances, with more threads becoming worthwhile at about $10$ seconds of sequential time.
Unfortunately, we even experience some minor slowdowns and the speedups are strongly scattered.
The maximum achieved speedups are
$\placeholder{flow_cutter}{max_speedup_flow_cutter_16}$, $\placeholder{flow_cutter}{max_speedup_flow_cutter_64}$
for \Partitioner{FlowCutter} and
$\placeholder{plain_flows}{max_speedup_plain_flows_16}$, $\placeholder{plain_flows}{max_speedup_plain_flows_64}$ for \Partitioner{ParPR-RL}.
These results match what we expected from~\cite{BaumstarkSyncPushRelabel}.
Restricted to instances with sequential running time $\geq 10$ seconds, the geometric mean speedups are
$\placeholder{flow_cutter}{gmean_speedup_flow_cutter_long_16}$ and $\placeholder{flow_cutter}{gmean_speedup_flow_cutter_long_64}$
for \Partitioner{FlowCutter} and
$\placeholder{plain_flows}{gmean_speedup_plain_flows_long_16}$, $\placeholder{plain_flows}{gmean_speedup_plain_flows_long_64}$ for \Partitioner{ParPR-RL}.

\begin{figure*}[!t]
  \begin{minipage}{.99\textwidth}
  \ifpdfplots
    \includegraphics{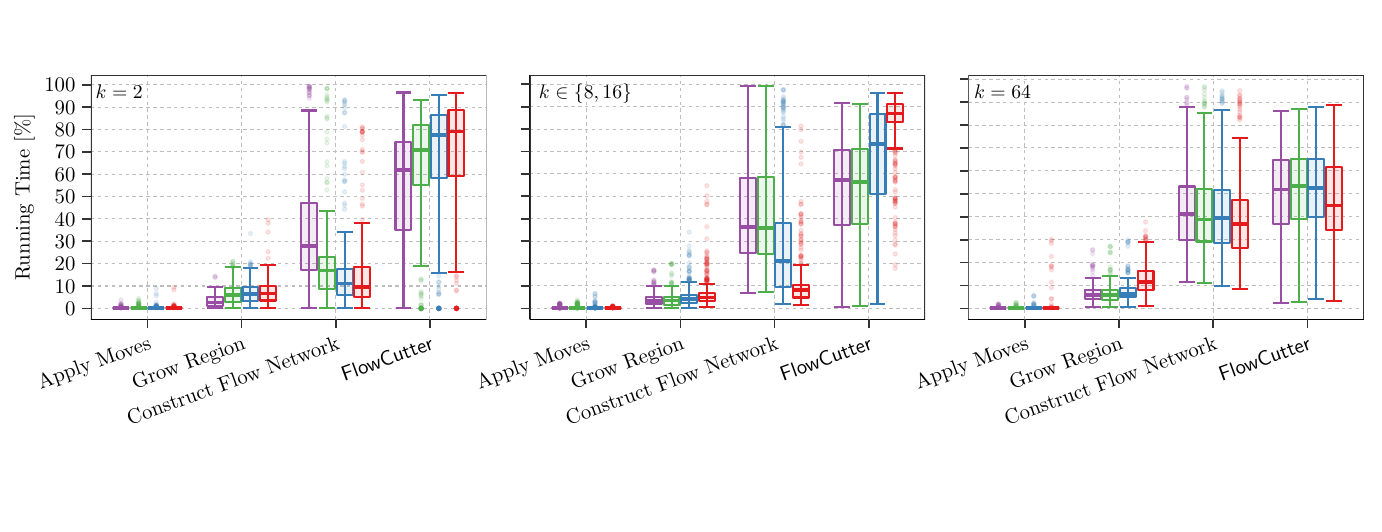}
  \fi
  \end{minipage} %
  \begin{minipage}{.99\textwidth}
    \vspace{-1.75cm}
    \centering
  \ifpdfplots
    \includegraphics{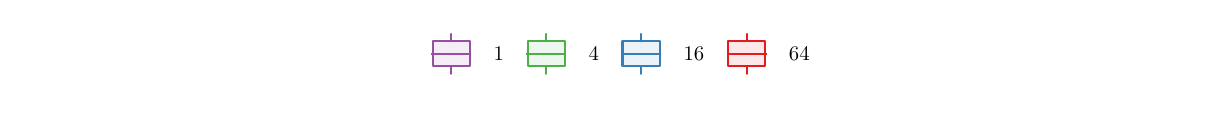}
  \fi
  \end{minipage} %
	\vspace{-1cm}
  \caption{Running times of the different phases of the flow-based refinement routine relative to its total running time
           for $k = 2$ (left), $k \in \{8,16\}$ (middle) and $k = 64$ (right) on set B.}
  \label{fig:running_time_shares}
\end{figure*}

\myparagraph{Mt-KaHyPar.}
We run the scalability experiments for \mtkahypardflows~on a subset of set B (76 out of 94 hypergraphs) that
contains all hypergraphs on which \mtkahypardflowsconfig{64} was able to complete in under $600$ seconds for
all $k \in \{2,8,16,64\}$\footnote{This experiment still took 4 weeks on machine B}.
We omit scalability experiments with \mtkahyparqflows~due to the long time requirements and because flow-based refinement is used in the same context in \mtkahypardflows.
Note that we use sequential implementations of the flow network construction and maximum flow algorithm in case
the number of flow problems processed in parallel is equal to the number of available threads.
Hence, scalability is limited by parallelization overheads and memory bandwidth, which makes achieving perfect speedups difficult.

The overall geometric mean speedup of \mtkahypardflows~is
$\placeholder{scalability}{gmean_speedup_mt_kahypar_4}$ for $t = 4$,
$\placeholder{scalability}{gmean_speedup_mt_kahypar_16}$ for $t = 16$ and
$\placeholder{scalability}{gmean_speedup_mt_kahypar_64}$ for $t = 64$.
If we only consider instances with a single-threaded running time $\ge 100$s, we achieve a geometric
mean speedup of $\placeholder{scalability}{gmean_speedup_large_mt_kahypar_64}$ for $t = 64$.

For $k = 2$, the scalability of the flow-based refinement routine largely depends on \Partitioner{FlowCutter} as the only parallelism source.
We can see that the speedups of the two are comparable (compare Figure~\ref{fig:speedups} top-middle with bottom-left).
There are a few outliers (e.g. \texttt{nlpkkt200} with a speedup of $\placeholder{scalability}{max_speedup_mt_kahypar_flow_64}$ for $t = 64$)
where the flow network construction dominates the overall execution time for $t = 1$.
For $k = 64$ and $t = 64$, we achieve a geometric mean speedup of $\placeholder{scalability}{gmean_speedup_mt_kahypar_flow_64_k64}$.
In this case, all parallelism is leveraged in the scheduler, and none in \Partitioner{FlowCutter}, which explains why we obtain
more reliable speedups than for all other $k$.
As the outer parallel construct, the scheduler is the more amenable parallelism source.
For $k \in \{8,16\}$, both parallelism sources are used.
The speedups are slightly better than for $k = 2$. Note that the poor speedups for instances with short single-threaded
running times ($\le 10$s) are caused by parallelization overheads of the network construction and maximum flow algorithm.

Figure~\ref{fig:running_time_shares} shows the running times of the different phases of the flow-based refinement routine
relative to its total running time. For $k \le 16$, \Partitioner{FlowCutter} dominates the running time.
For $k = 64$, the flow network construction and \Partitioner{FlowCutter} have the same share on the total running time, while applying move sequences and growing the region $B$ are negligible.

\begin{figure*}[!t]
  \begin{minipage}{.99\textwidth}
  \ifpdfplots
    \includegraphics{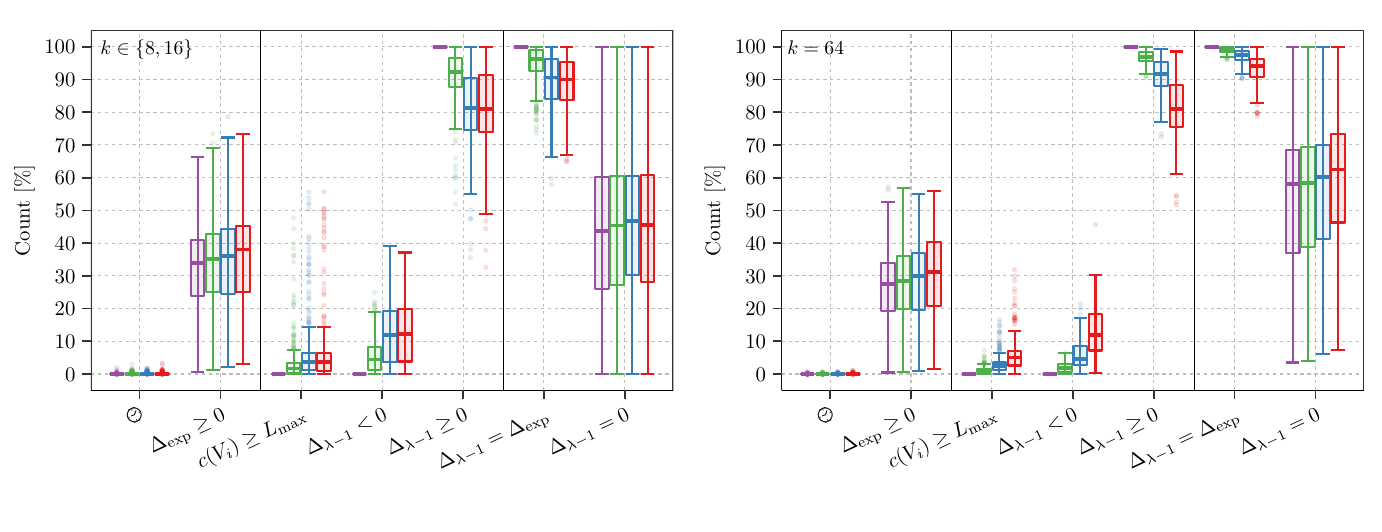}
  \fi
  \end{minipage} %
  \begin{minipage}{.99\textwidth}
    \vspace{-1.25cm}
    \centering
  \ifpdfplots
    \includegraphics{pdf_plots/flow_insights_stats_legend_set_b.pdf}
  \fi
  \end{minipage} %
	\vspace{-0.5cm}
  \caption{Conflicts for $k \in \{8,16\}$ (left) and $k = 64$ (right) on set B.
           For each instance, we count the refinements that exceed the time limit (\ClockLogo), the
           potential improvements ($\Delta_{\text{exp}} \ge 0$) and
           move sequences that violate the balance constraint ($c(V_i) \ge L_{\max}$) or degrade ($\Delta_{\lambda - 1} < 0$) or improve the
           connectivity metric ($\Delta_{\lambda - 1} \ge 0$). For move sequences with $\Delta_{\lambda - 1} \ge 0$, we
           count if the actual improvements equals the expected ($\Delta_{\lambda - 1} = \Delta_{\text{exp}}$)
           and zero-gain improvements ($\Delta_{\lambda - 1} = 0$).
       }
  \label{fig:conflict_rates}
\end{figure*}

\myparagraph{Search Interference.}
Figure~\ref{fig:conflict_rates} gives an overview on the different types of conflicts in the
flow-based refinement routine (as explained in Section~\ref{sec:scheduling}) and how often they occur.
In the median, $\placeholder{conflicts}{med_found_improvement_t64}\%$ of flow-based refinements find a potential
improvement, of which we successfully apply $\placeholder{conflicts}{med_success_t64}\%$ to the global partition for $t = 64$
($\placeholder{conflicts}{med_success_t16}\%$ for $t = 16$ and $\placeholder{conflicts}{med_success_t4}\%$ for $t = 4$).
For $t = 64$, $\placeholder{conflicts}{med_balance_violations_t64}\%$ of the move sequences violate the balance constraint
($\placeholder{conflicts}{med_balance_violations_t16}\%$ for $t = 16$ and $\placeholder{conflicts}{med_balance_violations_t4}\%$ for $t = 4$)
and $\placeholder{conflicts}{med_degrade_solution_t64}\%$ actually degrade the solution quality
($\placeholder{conflicts}{med_degrade_solution_t16}\%$ for $t = 16$ and $\placeholder{conflicts}{med_degrade_solution_t4}\%$ for $t = 4$).
However, increasing the number of threads does not adversely affect the solution quality of \mtkahypardflows~(see
Figure~\ref{fig:performance_profile_threads_set_b} in Appendx~\ref{appendix:quality_threads}).

\vspace{-0.25cm}

\section{Conclusion and Future Work}\label{sec:conclusion}
This work marks the end of a series of publications with the aim to transfer techniques used in
modern sequential partitioning algorithms into the shared-memory context without comprises in solution
quality. The result is a set of parallel algorithms unified in one framework
(\Partitioner{Mt-KaHyPar}) that outperforms all popular hypergraph partitioners~\cite{MT-KAHYPAR, MT-KAHYPAR-Q}.

Summarizing our experimental results, we obtain good speedups for medium size values of $k$, where we can rely on the scheduler, and acceptable speedups for small $k$, where the parallelism stems from the maximum flow algorithm.
Using 10 threads, our system is 10 times faster than the sequential state-of-the-art system \Partitioner{KaHyPar} with flow-based refinement, while achieving the same solution quality (connectivity metric).

Future work includes a deterministic version of parallel flow-based refinement to extend the existing deterministic algorithm~\cite{mt-kahypar-det-tr}, as well as
a highly localized version used in the $n$-level partitioner~\cite{MT-KAHYPAR-Q} that only constructs
small flow problems around uncontracted nodes.

\bibliography{tr}

\begin{thebibliography}{10}

\bibitem{KAHYPAR-K}
Yaroslav Akhremtsev, Tobias Heuer, Peter Sanders, and Sebastian Schlag.
\newblock {Engineering a Direct \emph{k}-way Hypergraph Partitioning
  Algorithm}.
\newblock In {\em 19th Workshop on Algorithm Engineering \& Experiments
  (ALENEX)}, pages 28--42. SIAM, 01 2017.
\newblock \href {https://doi.org/10.1137/1.9781611974768.3}
  {\path{doi:10.1137/1.9781611974768.3}}.

\bibitem{MT-KAHIP}
Yaroslav Akhremtsev, Peter Sanders, and Christian Schulz.
\newblock {High-Quality Shared-Memory Graph Partitioning}.
\newblock In {\em European Conference on Parallel Processing (Euro-Par)}, pages
  659--671. Springer, 8 2017.
\newblock \href {https://doi.org/10.1007/978-3-319-96983-1\_47}
  {\path{doi:10.1007/978-3-319-96983-1\_47}}.

\bibitem{ISPD98}
Charles~J. Alpert.
\newblock {The ISPD98 Circuit Benchmark Suite}.
\newblock In {\em International Symposium on Physical Design (ISPD)}, pages
  80--85, 4 1998.
\newblock \href {https://doi.org/10.1145/274535.274546}
  {\path{doi:10.1145/274535.274546}}.

\bibitem{ALPERT-SURVEY}
Charles~J. Alpert and Andrew~B. Kahng.
\newblock {Recent Directions in Netlist Partitioning: A Survey}.
\newblock {\em Integration}, 19(1-2):1--81, 1995.
\newblock \href {https://doi.org/10.1016/0167-9260(95)00008-4}
  {\path{doi:10.1016/0167-9260(95)00008-4}}.

\bibitem{andersen2008algorithm}
Reid Andersen. and Kevin~J. Lang.
\newblock {An Algorithm for Improving Graph Partitions}.
\newblock In {\em Proc. of the 19th ACM-SIAM Symposium on Discrete Algorithms},
  pages 651--660. Society for Industrial and Applied Mathematics, 2008.
\newblock \href {https://doi.org/10.5555/1347082.1347154}
  {\path{doi:10.5555/1347082.1347154}}.

\bibitem{AndersonSetubalPushRelabel}
Richard~J. Anderson and Jo{\~{a}}o~C. Setubal.
\newblock {A Parallel Implementation of the Push-Relabel Algorithm for the
  Maximum Flow Problem}.
\newblock {\em J. Parallel Distributed Comput.}, 29(1):17--26, 1995.
\newblock \href {https://doi.org/10.1006/jpdc.1995.1103}
  {\path{doi:10.1006/jpdc.1995.1103}}.

\bibitem{google-search}
Aaron Archer, Kevin Aydin, MohammadHossein Bateni, Vahab~S. Mirrokni, Aaron
  Schild, Ray Yang, and Richard Zhuang.
\newblock {Cache-Aware Load Balancing of Data Center Applications}.
\newblock {\em Proceedings of the {VLDB} Endowment}, 12(6):709--723, 2019.
\newblock \href {https://doi.org/10.14778/3311880.3311887}
  {\path{doi:10.14778/3311880.3311887}}.

\bibitem{kPaToH}
Cevdet Aykanat, Berkant~Barla Cambazoglu, and Bora U{\c{c}}ar.
\newblock {Multi-level Direct $k$-way Hypergraph Partitioning With Multiple
  Constraints and Fixed Vertices}.
\newblock {\em Journal of Parallel and Distributed Computing}, 68(5):609--625,
  2008.
\newblock \href {https://doi.org/10.1016/j.jpdc.2007.09.006}
  {\path{doi:10.1016/j.jpdc.2007.09.006}}.

\bibitem{GRAPH-SURVEY}
David~A. Bader, Henning Meyerhenke, Peter Sanders, and Dorothea Wagner.
\newblock {\em {Graph Partitioning and Graph Clustering}}, volume 588.
\newblock American Mathematical Society Providence, RI, 2013.
\newblock \href {https://doi.org/10.1090/conm/588}
  {\path{doi:10.1090/conm/588}}.

\bibitem{BaumstarkSyncPushRelabel}
Niklas Baumstark, Guy~E. Blelloch, and Julian Shun.
\newblock Efficient implementation of a synchronous parallel push-relabel
  algorithm.
\newblock In Nikhil Bansal and Irene Finocchi, editors, {\em Algorithms - {ESA}
  2015 - 23rd Annual European Symposium, Patras, Greece, September 14-16, 2015,
  Proceedings}, volume 9294 of {\em Lecture Notes in Computer Science}, pages
  106--117. Springer, 2015.
\newblock \href {https://doi.org/10.1007/978-3-662-48350-3\_10}
  {\path{doi:10.1007/978-3-662-48350-3\_10}}.

\bibitem{SAT14}
Anton Belov, Daniel Diepold, Marijn Heule, and Matti J{\"{a}}rvisalo.
\newblock {The SAT Competition 2014}.
\newblock \url{http://www.satcompetition.org/2014/}, 2014.

\bibitem{BUI}
Thang~N. Bui and Curt Jones.
\newblock {A Heuristic for Reducing Fill-In in Sparse Matrix Factorization}.
\newblock In {\em 6th {SIAM} Conference on Parallel Processing for Scientific
  Computing (PPSC)}, pages 445--452, 1993.
\newblock URL: \url{https://www.osti.gov/biblio/54439}.

\bibitem{PATOH}
Ümit~V. Catalyurek and Cevdet Aykanat.
\newblock {Hypergraph-Partitioning-based Decomposition for Parallel
  Sparse-Matrix Vector Multiplication}.
\newblock {\em IEEE Transactions on Parallel and Distributed Systems},
  10(7):673--693, 1999.
\newblock \href {https://doi.org/10.1109/71.780863}
  {\path{doi:10.1109/71.780863}}.

\bibitem{DBLP:journals/algorithmica/CherkasskyG97}
Boris~V. Cherkassky and Andrew~V. Goldberg.
\newblock {On Implementing the Push-Relabel Method for the Maximum Flow
  Problem}.
\newblock {\em Algorithmica}, 19(4):390--410, 1997.
\newblock \href {https://doi.org/10.1007/PL00009180}
  {\path{doi:10.1007/PL00009180}}.

\bibitem{schism}
Carlo Curino, Yang Zhang, Evan P.~C. Jones, and Samuel Madden.
\newblock {Schism: a Workload-Driven Approach to Database Replication and
  Partitioning}.
\newblock {\em Proceedings of the VLDB Endowment}, 3(1):48--57, 2010.
\newblock \href {https://doi.org/10.14778/1920841.1920853}
  {\path{doi:10.14778/1920841.1920853}}.

\bibitem{SPM}
Timothy~A. Davis and Yifan Hu.
\newblock {The University of Florida Sparse Matrix Collection}.
\newblock {\em ACM Transactions on Mathematical Software}, 38(1):1:1--1:25, 11
  2011.
\newblock \href {https://doi.org/10.1145/2049662.2049663}
  {\path{doi:10.1145/2049662.2049663}}.

\bibitem{delling2010graph}
Daniel Delling, Andrew~V. Goldberg, Ilya Razenshteyn, and Renato~F. Werneck.
\newblock {Graph Partitioning with Natural Cuts}.
\newblock In {\em Proc. of the 25th International Parallel and Distributed
  Processing Symposium}, pages 1135--1146, 2011.
\newblock \href {https://doi.org/10.1109/IPDPS.2011.108}
  {\path{doi:10.1109/IPDPS.2011.108}}.

\bibitem{INR}
Mehmet Deveci, Kamer Kaya, and {\"{U}}mit~V. {\c{C}}ataly{\"{u}}rek.
\newblock {Hypergraph Sparsification and Its Application to Partitioning}.
\newblock In {\em 42nd International Conference on Parallel Processing, {ICPP}
  2013, Lyon, France, October 1-4, 2013}, pages 200--209, 2013.
\newblock \href {https://doi.org/10.1109/ICPP.2013.29}
  {\path{doi:10.1109/ICPP.2013.29}}.

\bibitem{ZOLTAN}
Karen~D. Devine, Erik~G. Boman, Robert~T. Heaphy, Rob~H. Bisseling, and
  Ümit~V. Catalyurek.
\newblock {Parallel Hypergraph Partitioning for Scientific Computing}.
\newblock In {\em IEEE Transactions on Parallel and Distributed Systems}, pages
  10--pp. IEEE, 2006.
\newblock \href {https://doi.org/10.1109/IPDPS.2006.1639359}
  {\path{doi:10.1109/IPDPS.2006.1639359}}.

\bibitem{Dinitz}
Yefim Dinitz.
\newblock {Algorithm for Solution of a Problem of Maximum Flow in a Network
  with Power Estimation}.
\newblock {\em Soviet Mathematics-Doklady}, 11(5):1277--1280, September 1970.

\bibitem{PERFORMANCE-PROFILES}
Elizabeth~D. Dolan and Jorge~J. Mor{\'{e}}.
\newblock {Benchmarking Optimization Software with Performance Profiles}.
\newblock {\em Mathematical Programming}, 91(2):201--213, 2002.
\newblock \href {https://doi.org/10.1007/s101070100263}
  {\path{doi:10.1007/s101070100263}}.

\bibitem{FM}
Charles~M. Fiduccia and Robert~M. Mattheyses.
\newblock {A Linear-Time Heuristic for Improving Network Partitions}.
\newblock In {\em 19th Conference on Design Automation (DAC)}, pages 175--181,
  1982.
\newblock \href {https://doi.org/10.1145/800263.809204}
  {\path{doi:10.1145/800263.809204}}.

\bibitem{MINCUT-THEOREM}
Lester~Randolph Ford and Delbert~R Fulkerson.
\newblock {Maximal Flow through a Network}.
\newblock {\em Canadian Journal of Mathematics}, 8:399--404, 1956.
\newblock \href {https://doi.org/10.4153/CJM-1956-045-5}
  {\path{doi:10.4153/CJM-1956-045-5}}.

\bibitem{PUSH-RELABEL}
Andrew~V. Goldberg and Robert~Endre Tarjan.
\newblock {A New Approach to the Maximum-Flow Problem}.
\newblock {\em Journal of the {ACM}}, 35(4):921--940, 1988.
\newblock \href {https://doi.org/10.1145/48014.61051}
  {\path{doi:10.1145/48014.61051}}.

\bibitem{FlowsPHard}
Leslie~M. Goldschlager, Ralph~A. Shaw, and John Staples.
\newblock {The Maximum Flow Problem is Log Space Complete for {P}}.
\newblock {\em Theoretical Computer Science}, 21:105--111, 1982.
\newblock \href {https://doi.org/10.1016/0304-3975(82)90092-5}
  {\path{doi:10.1016/0304-3975(82)90092-5}}.

\bibitem{KAHYPAR-HFC}
Lars Gottesb{\"u}ren, Michael Hamann, Sebastian Schlag, and Dorothea Wagner.
\newblock {Advanced Flow-Based Multilevel Hypergraph Partitioning}.
\newblock {\em 18th International Symposium on Experimental Algorithms (SEA)},
  2020.
\newblock \href {https://doi.org/10.4230/LIPIcs.SEA.2020.11}
  {\path{doi:10.4230/LIPIcs.SEA.2020.11}}.

\bibitem{REBAHFC}
Lars Gottesb{\"u}ren, Michael Hamann, and Dorothea Wagner.
\newblock {Evaluation of a Flow-Based Hypergraph Bipartitioning Algorithm}.
\newblock In {\em 27th European Symposium on Algorithms (ESA)}, pages
  52:1--52:17, 2019.
\newblock \href {https://doi.org/10.4230/LIPIcs.ESA.2019.52}
  {\path{doi:10.4230/LIPIcs.ESA.2019.52}}.

\bibitem{MT-KAHYPAR-Q}
Lars Gottesb{\"{u}}ren, Tobias Heuer, Peter Sanders, and Sebastian Schlag.
\newblock {Shared-Memory $n$-level Hypergraph Partitioning}.
\newblock In {\em Workshop on Algorithm Engineering and Experiments (ALENEX)}.
  SIAM, 2022.
\newblock to appear.
\newblock URL: \url{https://arxiv.org/abs/2104.08107}.

\bibitem{mt-kahypar-det-tr}
Lars Gottesbüren and Michael Hamann.
\newblock Deterministic parallel hypergraph partitioning.
\newblock {\em CoRR}, abs/2112.12704, 2021.
\newblock URL: \url{https://arxiv.org/abs/2112.12704}, \href
  {http://arxiv.org/abs/2112.12704} {\path{arXiv:2112.12704}}.

\bibitem{MT-KAHYPAR}
{Gottesbüren, Lars and Heuer, Tobias and Sanders, Peter and Schlag,
  Sebastian}.
\newblock {Scalable Shared-Memory Hypergraph Partitioning}.
\newblock In {\em 23st Workshop on Algorithm Engineering \& Experiments
  (ALENEX)}. SIAM, 01 2021.
\newblock \href {https://doi.org/10.1137/1.9781611976472.2}
  {\path{doi:10.1137/1.9781611976472.2}}.

\bibitem{gray2021hyper}
Johnnie Gray and Stefanos Kourtis.
\newblock Hyper-optimized tensor network contraction.
\newblock {\em Quantum}, 5:410, 2021.
\newblock \href {https://doi.org/10.22331/q-2021-03-15-410}
  {\path{doi:10.22331/q-2021-03-15-410}}.

\bibitem{FLOW-CUTTER}
Michael Hamann and Ben Strasser.
\newblock {Graph Bisection with Pareto Optimization}.
\newblock {\em {ACM} Journal of Experimental Algorithmics}, 23, 2018.
\newblock \href {https://doi.org/10.1145/3173045} {\path{doi:10.1145/3173045}}.

\bibitem{KAHYPAR-MF}
Tobias Heuer, Peter Sanders, and Sebastian Schlag.
\newblock {Network Flow-Based Refinement for Multilevel Hypergraph
  Partitioning}.
\newblock {\em {ACM} Journal of Experimental Algorithmics (JEA)},
  24(1):2.3:1--2.3:36, 09 2019.
\newblock \href {https://doi.org/10.1145/3329872} {\path{doi:10.1145/3329872}}.

\bibitem{KAHYPAR-CA}
Tobias Heuer and Sebastian Schlag.
\newblock {Improving Coarsening Schemes for Hypergraph Partitioning by
  Exploiting Community Structure}.
\newblock In {\em 16th International Symposium on Experimental Algorithms
  (SEA)}, pages 21:1--21:19. Schloss Dagstuhl -- Leibniz-Zentrum f{\"u}r
  Informatik, 06 2017.
\newblock \href {https://doi.org/10.4230/LIPIcs.SEA.2017.21}
  {\path{doi:10.4230/LIPIcs.SEA.2017.21}}.

\bibitem{HongHeAsyncPushRelabel}
Bo~Hong and Zhengyu He.
\newblock {An Asynchronous Multithreaded Algorithm for the Maximum Network Flow
  Problem with Nonblocking Global Relabeling Heuristic}.
\newblock {\em {IEEE} Transaction on Parallel Distributed Systems},
  22(6):1025--1033, 2011.
\newblock \href {https://doi.org/10.1109/TPDS.2010.156}
  {\path{doi:10.1109/TPDS.2010.156}}.

\bibitem{SHP}
Igor Kabiljo, Brian Karrer, Mayank Pundir, Sergey Pupyrev, Alon Shalita,
  Yaroslav Akhremtsev, and Alessandro Presta.
\newblock {Social Hash Partitioner: A Scalable Distributed Hypergraph
  Partitioner}.
\newblock In {\em Proceedings of the VLDB Endowment}, volume~10, pages
  1418--1429, 2017.
\newblock \href {https://doi.org/10.14778/3137628.3137650}
  {\path{doi:10.14778/3137628.3137650}}.

\bibitem{ColoringPushRelabel}
G{\"{o}}k{\c{c}}ehan Kara and Can~C. {\"{O}}zturan.
\newblock {Graph Coloring Based Parallel Push-relabel Algorithm for the Maximum
  Flow Problem}.
\newblock {\em {ACM} Transactions on Mathematical Software}, 45(4):46:1--46:28,
  2019.
\newblock \href {https://doi.org/10.1145/3330481} {\path{doi:10.1145/3330481}}.

\bibitem{HMETIS}
George Karypis, Rajat Aggarwal, Vipin Kumar, and Shashi Shekhar.
\newblock {Multilevel Hypergraph Partitioning: Applications in VLSI Domain}.
\newblock {\em IEEE Transactions on Very Large Scale Integration (VLSI)
  Systems}, 7(1):69--79, 1999.
\newblock \href {https://doi.org/10.1109/92.748202}
  {\path{doi:10.1109/92.748202}}.

\bibitem{HMETIS-K}
George Karypis and Vipin Kumar.
\newblock {Multilevel \emph{k}-way Hypergraph Partitioning}.
\newblock {\em {VLSI} Design}, 2000(3):285--300, 2000.
\newblock \href {https://doi.org/10.1155/2000/19436}
  {\path{doi:10.1155/2000/19436}}.

\bibitem{PREFLOW}
Alexander~V. Karzanov.
\newblock {Determining the Maximal Flow in a Network by the Method of
  Preflows}.
\newblock In {\em Soviet Mathematics Doklady}, volume~15, pages 434--437, 1974.

\bibitem{KL}
Brian~W. Kernighan and Shen Lin.
\newblock {An Efficient Heuristic Procedure for Partitioning Graphs}.
\newblock {\em The Bell System Technical Journal}, 49(2):291--307, 2 1970.

\bibitem{DBLP:journals/tc/Krishnamurthy84}
Balakrishnan Krishnamurthy.
\newblock An improved min-cut algorithm for partitioning {VLSI} networks.
\newblock {\em {IEEE} Trans. Computers}, 33(5):438--446, 1984.
\newblock \href {https://doi.org/10.1109/TC.1984.1676460}
  {\path{doi:10.1109/TC.1984.1676460}}.

\bibitem{lang2004flow}
Kevin~J. Lang and Satish Rao.
\newblock {A Flow-Based Method for Improving the Expansion or Conductance of
  Graph Cuts}.
\newblock In {\em Proc. of 10th International Integer Programming and
  Combinatorial Optimization Conference}, volume 3064 of {\em LNCS}, pages
  383--400. Springer, 2004.
\newblock \href {https://doi.org/10.1007/978-3-540-25960-2\_25}
  {\path{doi:10.1007/978-3-540-25960-2\_25}}.

\bibitem{MT-METIS}
Dominique LaSalle and George Karypis.
\newblock {Multi-Threaded Graph Partitioning}.
\newblock In {\em IEEE Transactions on Parallel and Distributed Systems}, pages
  225--236. IEEE, 2013.
\newblock \href {https://doi.org/10.1109/IPDPS.2013.50}
  {\path{doi:10.1109/IPDPS.2013.50}}.

\bibitem{Lawler}
Eugene~L. Lawler.
\newblock {Cutsets and Partitions of Hypergraphs}.
\newblock {\em Networks}, 3(3):275--285, 1973.
\newblock \href {https://doi.org/10.1002/net.3230030306}
  {\path{doi:10.1002/net.3230030306}}.

\bibitem{LENGAUER}
Thomas Lengauer.
\newblock {\em {Combinatorial Algorithms for Integrated Circuit Layout}}.
\newblock John Wiley \& Sons, Inc., 1990.
\newblock \href {https://doi.org/10.1017/S0263574700015691}
  {\path{doi:10.1017/S0263574700015691}}.

\bibitem{BIPART}
Sepideh Maleki, Udit Agarwal, Martin Burtscher, and Keshav Pingali.
\newblock {BiPart: A Parallel and Deterministic Hypergraph Partitioner}.
\newblock In {\em Proceedings of the 26th ACM SIGPLAN Symposium on Principles
  and Practice of Parallel Programming}, pages 161--174, 2021.
\newblock \href {https://doi.org/10.1145/3437801.3441611}
  {\path{doi:10.1145/3437801.3441611}}.

\bibitem{MANN-PAPA14}
Zoltán~{\'{A}}. Mann and Pál~A. Papp.
\newblock {Formula Partitioning Revisited}.
\newblock In {\em 5th Pragmatics of {SAT} Workshop}, pages 41--56, 2014.
\newblock \href {https://doi.org/10.29007/9skn} {\path{doi:10.29007/9skn}}.

\bibitem{ADAPTIVE-STOP-RULE}
Vitaly Osipov and Peter Sanders.
\newblock {n-Level Graph Partitioning}.
\newblock In {\em 18th European Symposium on Algorithms (ESA)}, pages 278--289.
  Springer, 2010.
\newblock \href {https://doi.org/10.1007/978-3-642-15775-2_24}
  {\path{doi:10.1007/978-3-642-15775-2_24}}.

\bibitem{PAPA-MARKOV}
David~A. Papa and Igor~L. Markov.
\newblock {Hypergraph Partitioning and Clustering}.
\newblock In {\em Handbook of Approximation Algorithms and Metaheuristics}.
  2007.
\newblock \href {https://doi.org/10.1201/9781420010749.ch61}
  {\path{doi:10.1201/9781420010749.ch61}}.

\bibitem{TBB}
Chuck Pheatt.
\newblock {Intel Threading Building Blocks}.
\newblock {\em Journal of Computing Sciences in Colleges}, 23(4):298--298,
  2008.

\bibitem{PicardQ82}
Jean{-}Claude Picard and Maurice Queyranne.
\newblock {On the Structure of All Minimum Cuts in a Network and Applications}.
\newblock {\em Math. Program.}, 22(1):121, 1982.
\newblock \href {https://doi.org/10.1007/BF01581031}
  {\path{doi:10.1007/BF01581031}}.

\bibitem{LABEL_PROPAGATION}
Usha~Nandini Raghavan, R{\'e}ka Albert, and Soundar Kumara.
\newblock {Near Linear Time Algorithm to Detect Community Structures in
  Large-Scale Networks}.
\newblock {\em Physical Review E}, 76(3):036106, 2007.
\newblock \href {https://doi.org/10.1103/PhysRevE.76.036106}
  {\path{doi:10.1103/PhysRevE.76.036106}}.

\bibitem{KAFFPA}
Peter Sanders and Christian Schulz.
\newblock {Engineering Multilevel Graph Partitioning Algorithms}.
\newblock In {\em 19th European Symposium on Algorithms (ESA)}, pages 469--480.
  Springer, 2011.
\newblock \href {https://doi.org/10.1007/978-3-642-23719-5_40}
  {\path{doi:10.1007/978-3-642-23719-5_40}}.

\bibitem{KAHYPAR-DIS}
Sebastian Schlag.
\newblock {High-Quality Hypergraph Partitioning}.
\newblock 2020.
\newblock \href {https://doi.org/10.5445/IR/1000105953}
  {\path{doi:10.5445/IR/1000105953}}.

\bibitem{KaHyPar-R}
Sebastian Schlag, Vitali Henne, Tobias Heuer, Henning Meyerhenke, Peter
  Sanders, and Christian Schulz.
\newblock {$k$-way Hypergraph Partitioning via n-Level Recursive Bisection}.
\newblock In {\em 18th Workshop on Algorithm Engineering \& Experiments
  (ALENEX)}, pages 53--67. SIAM, 2016.
\newblock \href {https://doi.org/10.1137/1.9781611974317.5}
  {\path{doi:10.1137/1.9781611974317.5}}.

\bibitem{ShiloachVishkin}
Yossi Shiloach and Uzi Vishkin.
\newblock {An O(n{\({^2}\)} log n) Parallel Max-Flow Algorithm}.
\newblock {\em Journal of Algorithms}, 3(2):128--146, 1982.
\newblock \href {https://doi.org/10.1016/0196-6774(82)90013-X}
  {\path{doi:10.1016/0196-6774(82)90013-X}}.

\bibitem{PARKWAY-2}
Aleksandar Trifunovic and William~J. Knottenbelt.
\newblock {Parkway 2.0: A Parallel Multilevel Hypergraph Partitioning Tool}.
\newblock In {\em International Symposium on Computer and Information
  Sciences}, pages 789--800. Springer, 2004.
\newblock \href {https://doi.org/10.1007/978-3-540-30182-0\_79}
  {\path{doi:10.1007/978-3-540-30182-0\_79}}.

\bibitem{DAC}
Natarajan Viswanathan, Charles~J. Alpert, Cliff C.~N. Sze, Zhuo Li, and
  Yaoguang Wei.
\newblock {The DAC 2012 Routability-Driven Placement Contest and Benchmark
  Suite}.
\newblock In {\em 49th Conference on Design Automation (DAC)}, pages 774--782.
  ACM, 6 2012.
\newblock \href {https://doi.org/10.1145/2228360.2228500}
  {\path{doi:10.1145/2228360.2228500}}.

\bibitem{yang-wong-fbb}
Hannah~Honghua Yang and D.F. Wong.
\newblock {Efficient Network Flow Based Min-Cut Balanced Partitioning}.
\newblock {\em IEEE Transactions on Computer-Aided Design of Integrated
  Circuits and Systems}, 15(12):1533--1540, 1996.
\newblock \href {https://doi.org/10.1007/978-1-4615-0292-0_41}
  {\path{doi:10.1007/978-1-4615-0292-0_41}}.

\end{thebibliography}

\clearpage

\appendix

\begin{filecontents*}{data/bfs_distance.dat}
gmean_delta_1 = 0.99
gmean_delta_2 = 1.24
gmean_delta_4 = 1.46
gmean_delta_8 = 1.6
gmean_delta_inf = 1.65
slower_delta_4 = 17.85
slower_delta_8 = 29.27
slower_delta_inf = 33.37
\end{filecontents*}
\DTLloaddb[noheader, keys={key,value}]{bfs_distance}{data/bfs_distance.dat}
\begin{filecontents*}{data/time_limit.dat}
gmean_kappa_2 = 1.22
gmean_kappa_4 = 1.23
gmean_kappa_8 = 1.24
gmean_kappa_16 = 1.24
gmean_kappa_inf = 1.26
\end{filecontents*}
\DTLloaddb[noheader, keys={key,value}]{time_limit}{data/time_limit.dat}
\begin{filecontents*}{data/parallel_search_mult.dat}
gmean_tau_05 = 14.63
gmean_tau_1 = 12.35
gmean_tau_2 = 12.87
gmean_tau_4 = 13.34
gmean_tau_max = 13.01
\end{filecontents*}
\DTLloaddb[noheader, keys={key,value}]{parallel_search_mult}{data/parallel_search_mult.dat}
\begin{filecontents*}{data/bulk_piercing.dat}
default_with_bulk = 2.73
default_without_bulk = 2.96
quality_with_bulk = 5.08
quality_without_bulk = 5.25
\end{filecontents*}
\DTLloaddb[noheader, keys={key,value}]{bulk_piercing}{data/bulk_piercing.dat}

\section{Parameter Tuning}\label{appendix:parameter_tuning}

The flow-based refinement algorithm has several parameters whose choice influences
the solution quality and running time of \Partitioner{Mt-KaHyPar}.
This section summarizes our parameter tuning experiments and explains the tradeoffs
that led to the final parameter configuration.
We summarize the running times of all evaluated configurations in Table~\ref{tbl:gmean_time_parameter_tuning}.

\begin{figure*}[!htb]
  \begin{minipage}{.99\textwidth}
  \ifpdfplots
    \includegraphics{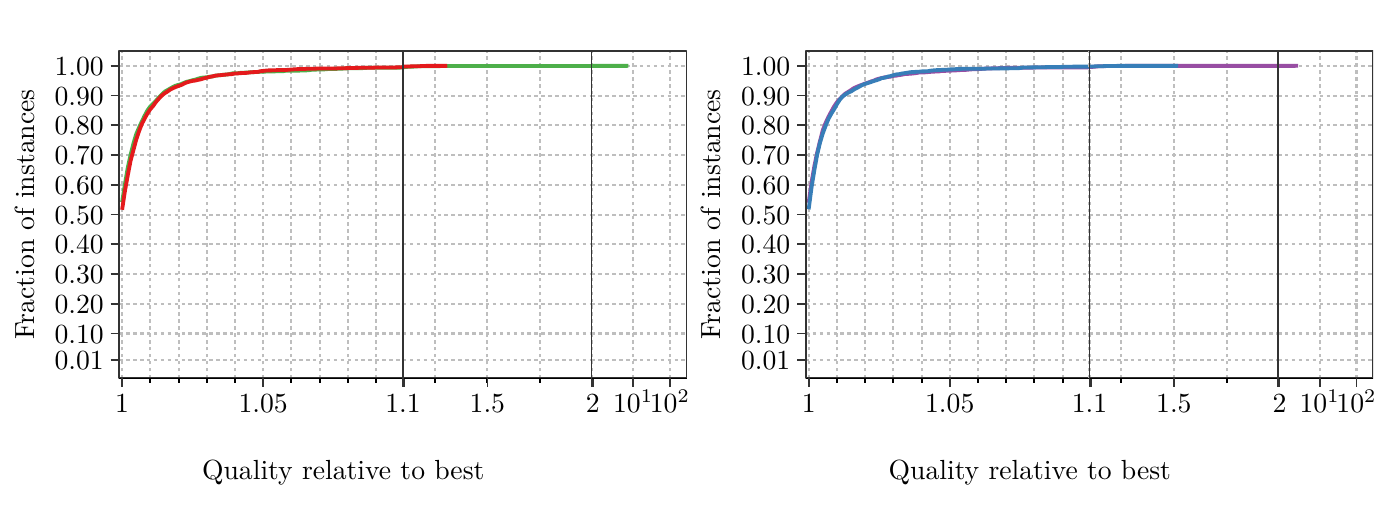}
  \fi
  \end{minipage} %
  \begin{minipage}{.99\textwidth}
    \vspace{-0.75cm}
    \centering
  \ifpdfplots
    \includegraphics{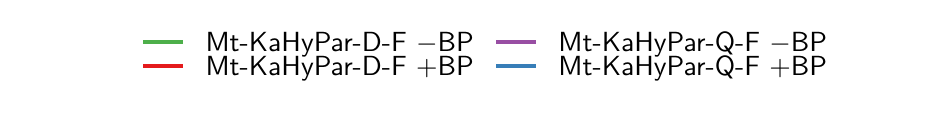}
  \fi
  \end{minipage} %
	\vspace{-0.5cm}
  \caption{Performance profiles comparing solution quality of \mtkahypardflows~and \mtkahyparqflows~with
           ($+$BP) and without bulk piercing ($-$BP) on set A.}
  \label{fig:performance_profile_bulk_piercing}
\end{figure*}

\myparagraph{Bulk Piercing.}
We evaluated the performance of the \Partitioner{FlowCutter} algorithm as flow-based refinement in \Partitioner{Mt-KaHyPar}
with (\Partitioner{$+$BP}) and without bulk piercing (\Partitioner{$-$BP}).
The experiments are conducted on set A and machine A with $k \in \{2,4,8,16,32,64,128\}$, $\varepsilon = 0.03$ and $10$ repetitions.
All configurations use $10$ threads.

Figure~\ref{fig:performance_profile_bulk_piercing} shows that there are no noticable differences in solution quality
when using bulk piercing. Furthermore, both variants that uses bulk piercing are slightly faster than their
counterparts that only pierces one vertex in each iteration (gmean time $\placeholder{bulk_piercing}{default_with_bulk}$s vs
$\placeholder{bulk_piercing}{default_without_bulk}$s, \mtkahypardflows, and $\placeholder{bulk_piercing}{quality_with_bulk}$s vs
$\placeholder{bulk_piercing}{quality_without_bulk}$s, \mtkahyparqflows). Thus, we enable bulk piercing per default.

\begin{figure*}[!htb]
  \begin{minipage}{.99\textwidth}
  \ifpdfplots
    \includegraphics{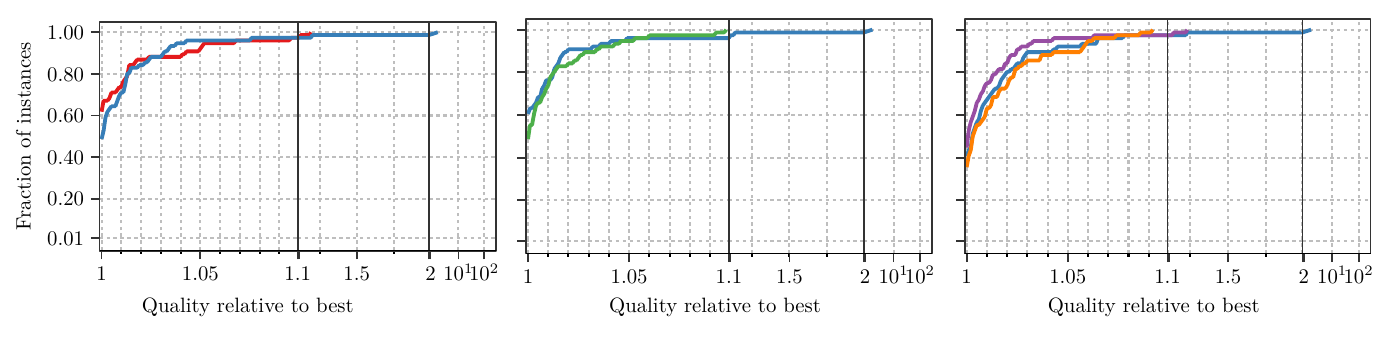}
  \fi
  \end{minipage} %
  \begin{minipage}{.99\textwidth}
    \vspace{-0.6cm}
    \centering
  \ifpdfplots
    \includegraphics{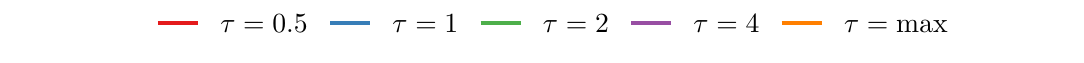}
  \fi
  \end{minipage} %
	\vspace{-0.5cm}
  \caption{Performance profiles comparing the solution quality for different values of $\tau$ that controls
           the available parallelism in our scheduler which is initialized with $\min(\min(t, \frac{k(k-1)}{2}),\tau \cdot k)$ threads.}
  \label{fig:parallel_search_mult}
\end{figure*}

\myparagraph{Scheduler Parallelism.} The scheduler starts $\min(\min(t, \frac{k(k-1)}{2}),\tau \cdot k)$ threads which
process the block pairs of the quotient graph in parallel.
We evaluated \mtkahypardflows~with $\tau \in \{0.5, 1, 2, 4, \infty\}$, $k \in \{2,8,16,64\}$, $\varepsilon = 0.03$ and $3$ repetitions
on a subset of set B\footnote{19 out of 94 instances: 5 VLSI, 5 SPM and 9 SAT instances.} and machine B. All configurations use $64$ threads.

\begin{figure*}[!htb]
  \begin{minipage}{.99\textwidth}
    \centering
  \ifpdfplots
    \includegraphics{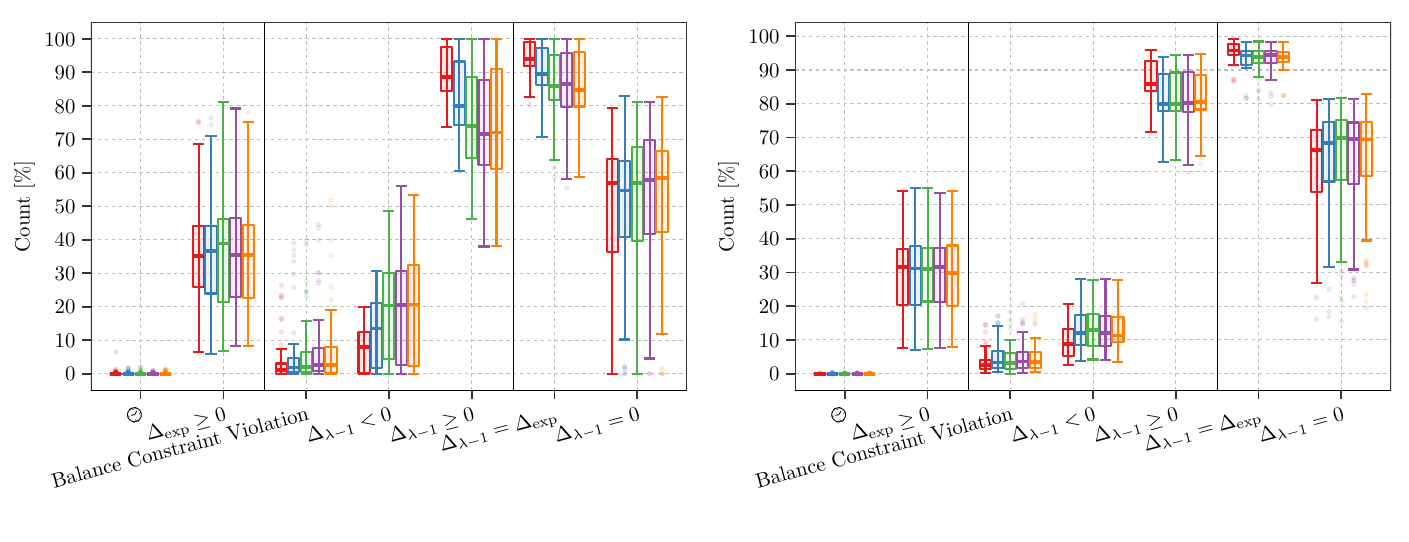}
  \fi
  \end{minipage} %
  \begin{minipage}{.99\textwidth}
    \vspace{-0.75cm}
    \centering
  \ifpdfplots
    \includegraphics{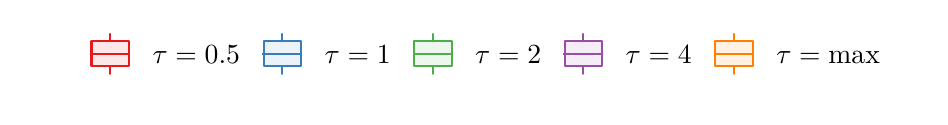}
  \fi
  \end{minipage} %
	\vspace{-0.55cm}
  \caption{Conflict rates for different values of $\tau$ with $k \in \{8,16\}$ (left) and $k = 64$ (right) on a subset of set B.
           The ticks are explained in Figure~\ref{fig:conflict_rates}.}
  \label{fig:conflicts_parallel_search_mult}
\end{figure*}

Figure~\ref{fig:parallel_search_mult} shows that all evaluated configurations for $\tau$ produces partitions with
comparable solution quality.
\mtkahypardflows~with $\tau = 1$ is faster than all other configurations.
Figure~\ref{fig:conflicts_parallel_search_mult} shows the number of interferences between
the threads for increasing values of $\tau$ in percentage.
We can see that the number of conflicts significantly increases for $\tau \ge 1$.
This also explains that the running times slightly increase for $\tau > 1$. Applying the move sequences
fails more often for larger values of $\tau$, which slows down the convergence
of the scheduling algorithm. Therefore, we choose $\tau = 1$.

\begin{figure*}[!htb]
  \begin{minipage}{.99\textwidth}
  \ifpdfplots
    \includegraphics{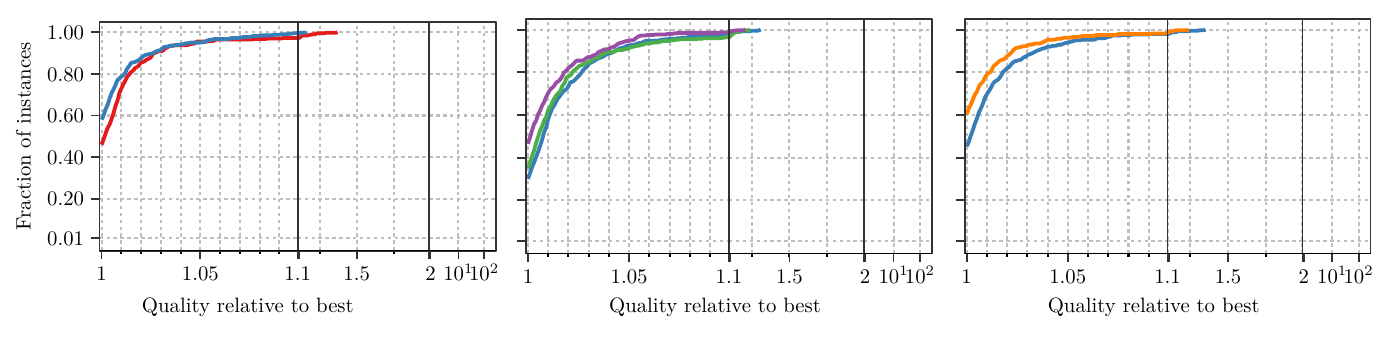}
  \fi
  \end{minipage} %
  \begin{minipage}{.99\textwidth}
    \vspace{-0.6cm}
    \centering
  \ifpdfplots
    \includegraphics{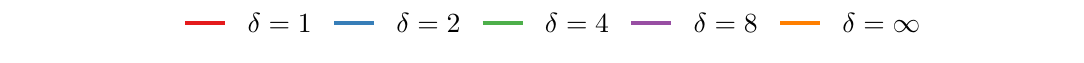}
  \fi
  \end{minipage} %
	\vspace{-0.5cm}
  \caption{Performance profiles comparing the solution quality if we restict
           the distance of each node to the cut hyperedges to be smaller or equal to $\delta$
           when we grow the region $B$.}
  \label{fig:bfs_distance}
\end{figure*}

\myparagraph{Maximum BFS Distance.} If we grow the region $B$ (defines the flow network)
around the cut hyperedges of a bipartition, we restrict the distance of each node to the
cut hyperedges to be smaller than or equal to $\delta$. We evaluate \mtkahypardflows~with
$\delta \in \{1,2,4,8,\infty\}$, $k \in \{2,8,16,64\}$, $\varepsilon = 0.03$ and $3$ repetitions
on a subset of set A\footnote{subset includes 100 instances which are also used in
the parameter tuning experiments of \Partitioner{KaHyPar}~\cite{KAHYPAR-DIS}.} and machine B.
All configurations use $16$ threads.

Figure~\ref{fig:bfs_distance} shows that the solution quality of \mtkahypardflows~with $\delta = 2$
is slightly better compared to those with $\delta = 1$. All configurations of \mtkahypardflows~with $\delta > 2$ produces partitions
that are marginally better than those with $\delta = 2$. However, \mtkahypardflows~is
$\placeholder{bfs_distance}{slower_delta_4}\%$ slower with $\delta = 4$ than with $\delta = 2$.
Therefore, we choose $\delta = 2$.


\begin{table*}[!htb]
  \centering
  \caption{Summarizes geometric mean running times of all configurations
           evaluated in our parameter tuning experiments.}
  \label{tbl:gmean_time_parameter_tuning}
  \vspace{-0.25cm}
  \begin{tabular}{lr|lr|lr}
   & $t[s]$ & & $t[s]$ &  & $t[s]$  \\
  \midrule
    $\delta = 1$                      & $\placeholder{bfs_distance}{gmean_delta_1}$ &
    $\tau = 0.5$                      & $\placeholder{parallel_search_mult}{gmean_tau_05}$ &
    \mtkahypardflowsconfig{$-$BP}     & $\placeholder{bulk_piercing}{default_without_bulk}$ \\
    $\delta = 2$                      & $\placeholder{bfs_distance}{gmean_delta_2}$ &
    $\tau = 1$                        & $\placeholder{parallel_search_mult}{gmean_tau_1}$ &
    \mtkahypardflowsconfig{$+$BP}     & $\placeholder{bulk_piercing}{default_with_bulk}$ \\
    $\delta = 4$                      & $\placeholder{bfs_distance}{gmean_delta_4}$ &
    $\tau = 2$                        & $\placeholder{parallel_search_mult}{gmean_tau_2}$ &
    \mtkahyparqflowsconfig{$-$BP}     & $\placeholder{bulk_piercing}{quality_without_bulk}$ \\
    $\delta = 8$                      & $\placeholder{bfs_distance}{gmean_delta_8}$ &
    $\tau = 4$                        & $\placeholder{parallel_search_mult}{gmean_tau_4}$ &
    \mtkahyparqflowsconfig{$+$BP}     & $\placeholder{bulk_piercing}{quality_with_bulk}$ \\
    $\delta = \infty$                 & $\placeholder{bfs_distance}{gmean_delta_inf}$ &
    $\tau = \max$                     & $\placeholder{parallel_search_mult}{gmean_tau_max}$ \\
  \end{tabular}
\end{table*}

\FloatBarrier

\section{Pairwise Comparisons with other Algorithms}\label{appendix:pairwise_comparisons}

\begin{figure*}[!htb]
  \begin{minipage}{.99\textwidth}
  \ifpdfplots
    \includegraphics{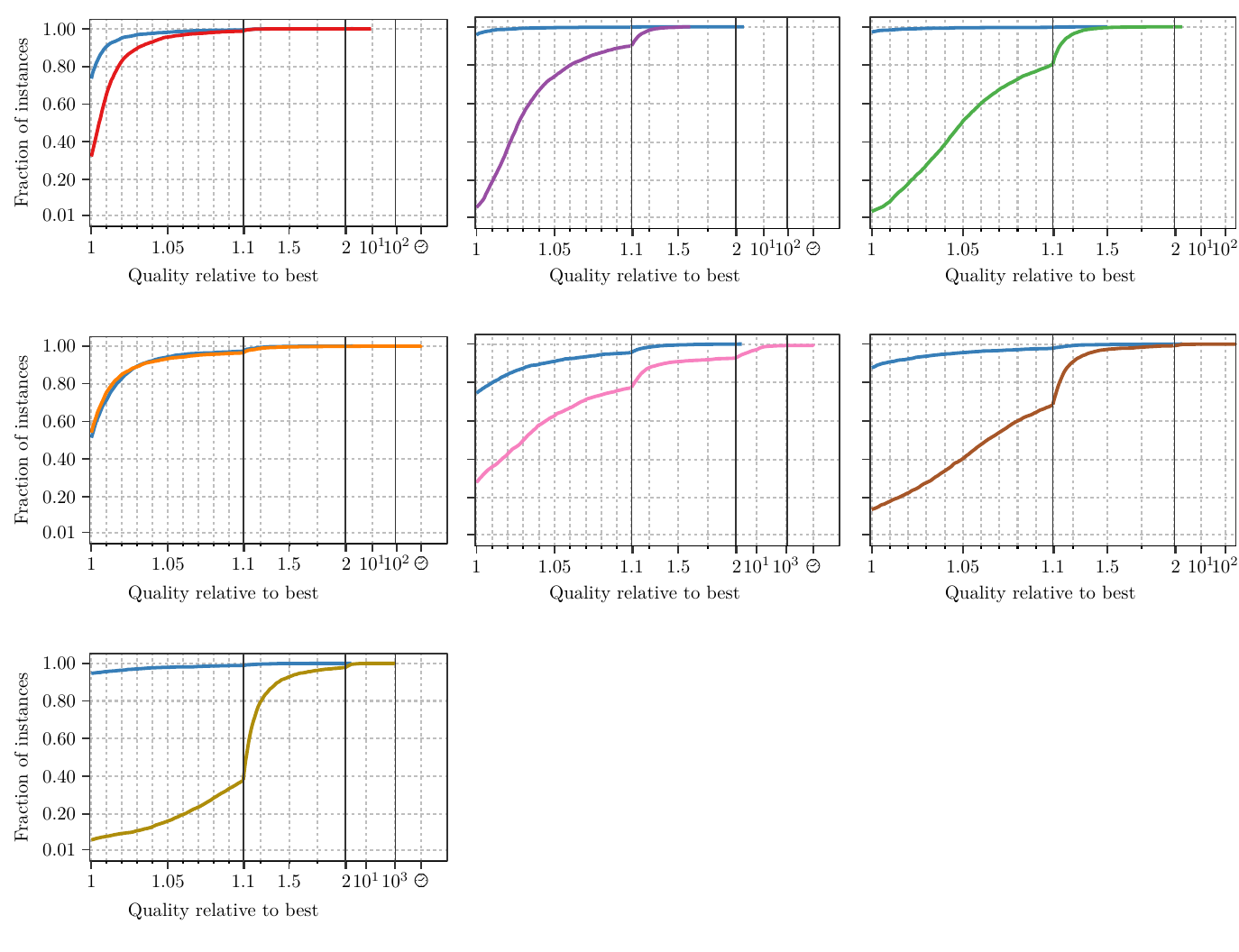}
  \fi
  \end{minipage} %
  \begin{minipage}{.99\textwidth}
    \vspace{-0.45cm}
    \centering
  \ifpdfplots
    \includegraphics{pdf_plots/performance_profile_set_a_legend.pdf}
  \fi
  \end{minipage} %
	\vspace{-0.25cm}
  \caption{Performance profiles comparing solution quality of \mtkahyparqflows~with different partitioners individually on set A.}
  \label{fig:quality_pairwise_set_a}
\end{figure*}

\begin{figure*}[!htb]
  \begin{minipage}{.99\textwidth}
  \ifpdfplots
    \includegraphics{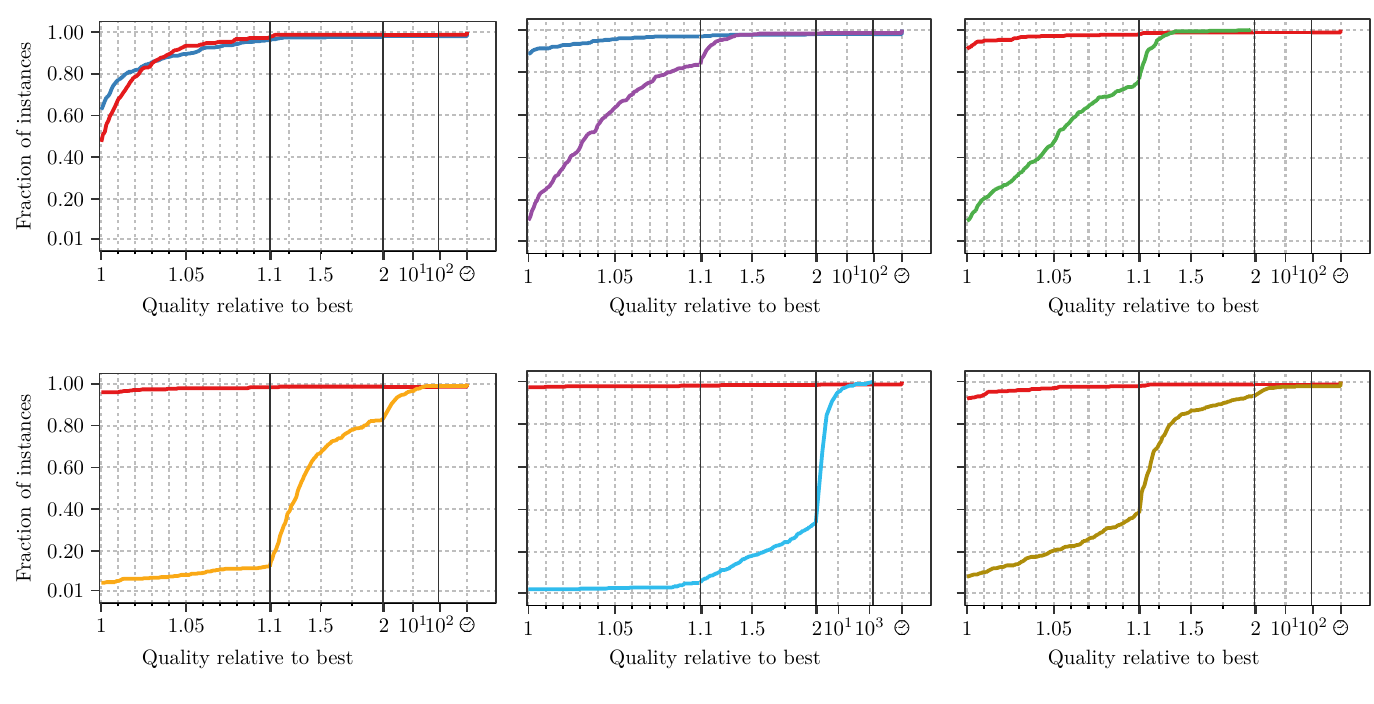}
  \fi
  \end{minipage} %
  \begin{minipage}{.99\textwidth}
    \vspace{-0.45cm}
    \centering
  \ifpdfplots
    \includegraphics{pdf_plots/performance_profile_set_b_legend.pdf}
  \fi
  \end{minipage} %
	\vspace{-0.25cm}
  \caption{Performance profiles comparing solution quality of \mtkahyparqflows~with different partitioners individually on set B.}
  \label{fig:quality_pairwise_set_b}
\end{figure*}

\FloatBarrier

\section{Quality with Increasing Number of Threads}\label{appendix:quality_threads}

\begin{figure*}[!htb]
  \begin{minipage}{.99\textwidth}
  \ifpdfplots
    \includegraphics{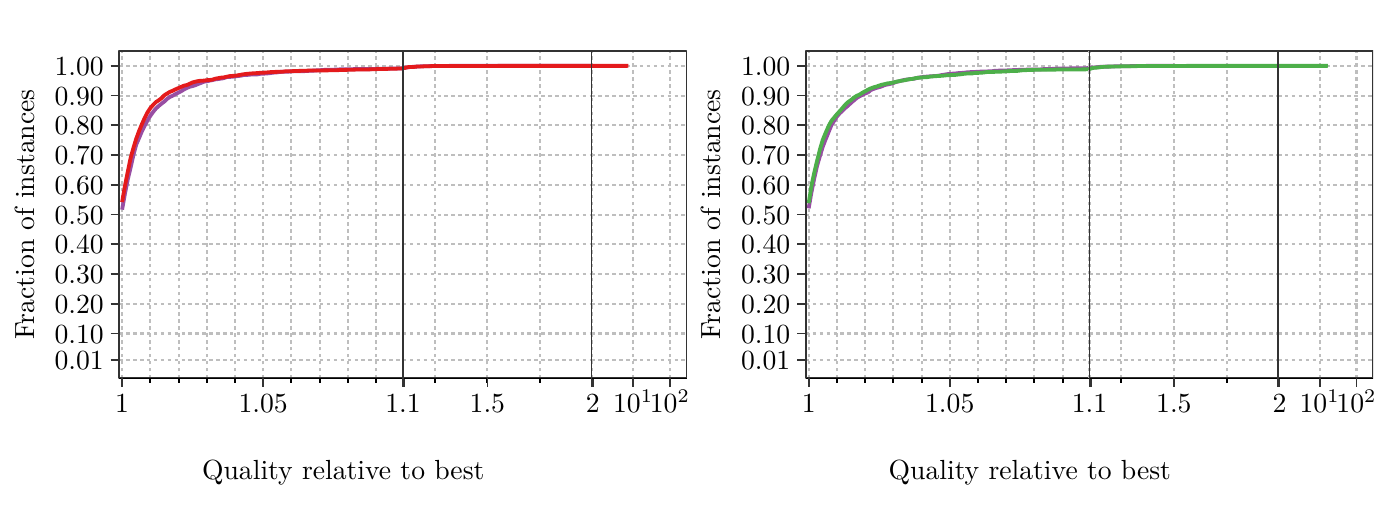}
  \fi
  \end{minipage} %
  \begin{minipage}{.99\textwidth}
    \vspace{-0.75cm}
    \centering
  \ifpdfplots
    \includegraphics{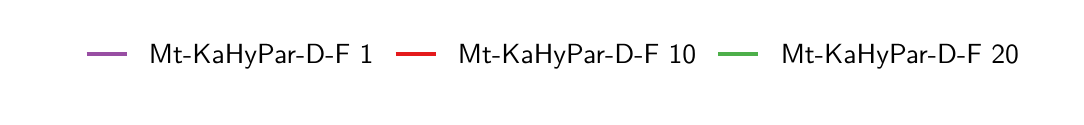}
  \fi
  \end{minipage} %
	\vspace{-0.5cm}
  \caption{Performance profiles comparing solution quality of \mtkahypardflows~with increasing number of threads on set A.}
\end{figure*}

\begin{figure*}[!htb]
  \begin{minipage}{.99\textwidth}
  \ifpdfplots
    \includegraphics{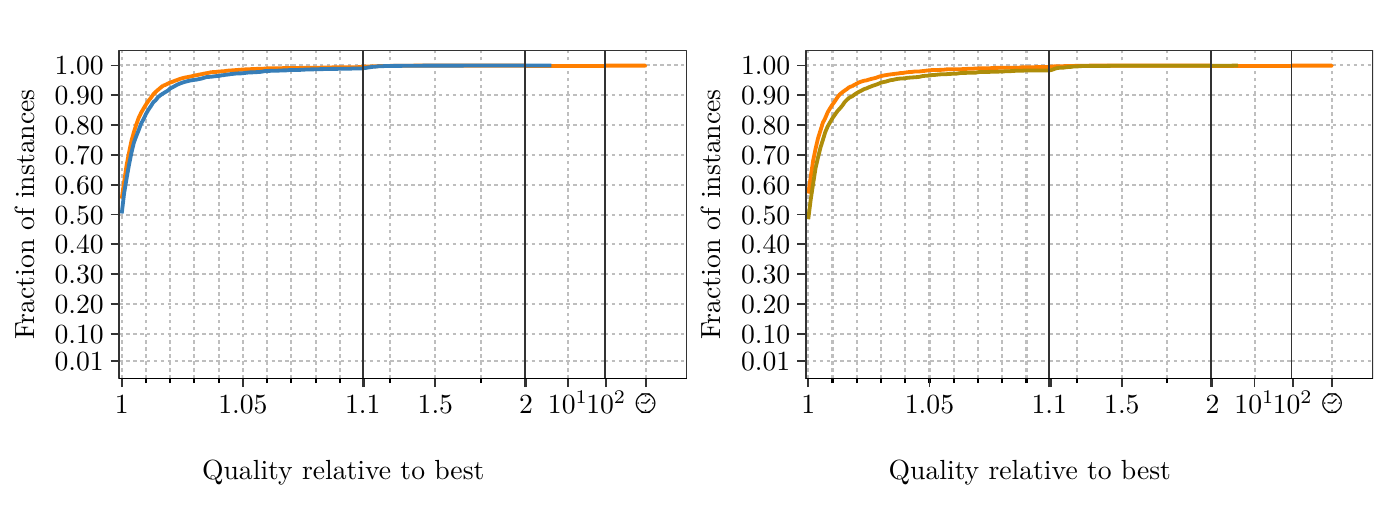}
  \fi
  \end{minipage} %
  \begin{minipage}{.99\textwidth}
    \vspace{-0.75cm}
    \centering
  \ifpdfplots
    \includegraphics{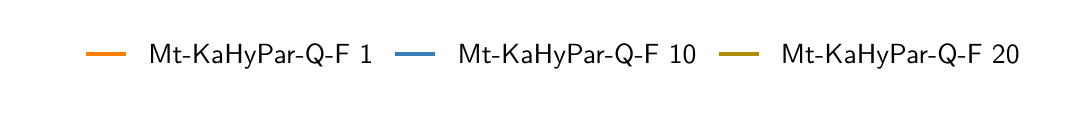}
  \fi
  \end{minipage} %
	\vspace{-0.5cm}
  \caption{Performance profiles comparing solution quality of \mtkahyparqflows~with increasing number of threads on set A.}
\end{figure*}

\begin{figure*}[!htb]
  \begin{minipage}{.99\textwidth}
  \ifpdfplots
    \includegraphics{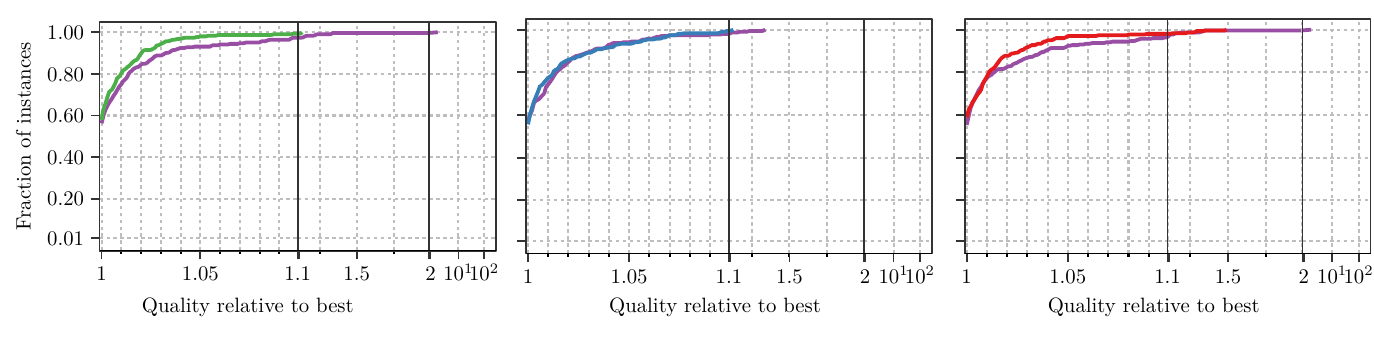}
  \fi
  \end{minipage} %
  \begin{minipage}{.99\textwidth}
    \vspace{-0.75cm}
    \centering
  \ifpdfplots
    \includegraphics{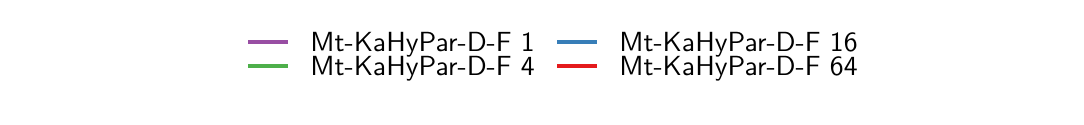}
  \fi
  \end{minipage} %
	\vspace{-0.5cm}
  \caption{Performance profiles comparing solution quality of \mtkahypardflows~with increasing number of threads on set B.}
  \label{fig:performance_profile_threads_set_b}
\end{figure*}

\begin{filecontents*}{data/set_a_scalability.dat}
gmean_mt_kahypar_hd_1 = 15.15
gmean_mt_kahypar_hd_10 = 2.73
gmean_mt_kahypar_hd_20 = 2.24
gmean_mt_kahypar_hq_1 = 34.3
gmean_mt_kahypar_hq_10 = 5.08
gmean_mt_kahypar_hq_20 = 3.78
\end{filecontents*}
\DTLloaddb[noheader, keys={key,value}]{scalability_set_a}{data/set_a_scalability.dat}

\begin{table*}[!htb]
  \centering
  \caption{Summarizes geometric mean running times of \Partitioner{Mt-KaHyPar}
           with increasing number of threads on set A and B.}
  \label{tbl:gmean_time_mt_kahypar}
  \vspace{-0.25cm}
  \begin{tabular}{lr|lr|lr}
  \multicolumn{4}{c|}{Set A} & \multicolumn{2}{c}{Set B} \\
  Algorithm & $t[s]$ & Algorithm & $t[s]$ & Algorithm & $t[s]$  \\
  \midrule
    \mtkahyparqflowsconfig{1}      & $\placeholder{scalability_set_a}{gmean_mt_kahypar_hq_1}$ &
    \mtkahypardflowsconfig{1}      & $\placeholder{scalability_set_a}{gmean_mt_kahypar_hd_1}$ &
    \mtkahypardflowsconfig{1}      & $\placeholder{scalability}{gmean_mt_kahypar_hd_1}$ \\
    \mtkahyparqflowsconfig{10}     & $\placeholder{scalability_set_a}{gmean_mt_kahypar_hq_10}$ &
    \mtkahypardflowsconfig{10}     & $\placeholder{scalability_set_a}{gmean_mt_kahypar_hd_10}$ &
    \mtkahypardflowsconfig{4}      & $\placeholder{scalability}{gmean_mt_kahypar_hd_4}$ \\
    \mtkahyparqflowsconfig{20}     & $\placeholder{scalability_set_a}{gmean_mt_kahypar_hq_20}$ &
    \mtkahypardflowsconfig{20}     & $\placeholder{scalability_set_a}{gmean_mt_kahypar_hd_20}$ &
    \mtkahypardflowsconfig{16}     & $\placeholder{scalability}{gmean_mt_kahypar_hd_16}$ \\
                                   &  &
                                   &  &
    \mtkahypardflowsconfig{64}     & $\placeholder{scalability}{gmean_mt_kahypar_hd_64}$ \\
  \end{tabular}
\end{table*}

\FloatBarrier

\section{Flow Algorithms}\label{appendix:flow_algo_comparison}

\begin{figure*}[!htb]
  \centering
  \ifpdfplots
    \includegraphics{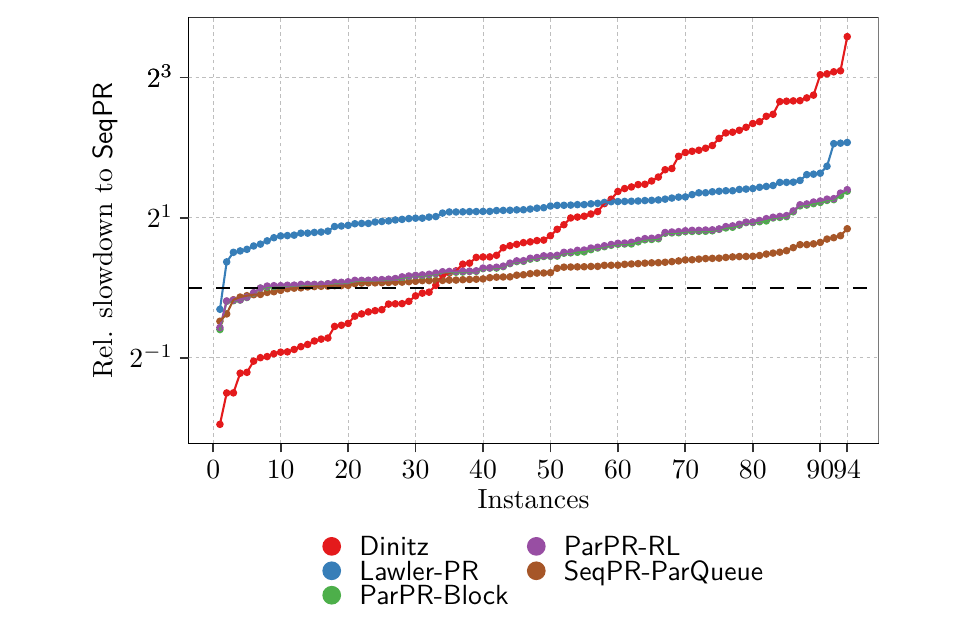}
  \fi
	\vspace{-0.25cm}
  \caption{Single-threaded running times of the evaluated flow algorithms relative to a sequential push-relabel implementation (\Partitioner{SeqPR}).}\label{fig:runtime:plain_flows}
\end{figure*}

In Figure~\ref{fig:runtime:plain_flows} we plot the running time of different flow algorithms relative to a sequential hypergraph-based FIFO push-relabel implementation (\Partitioner{SeqPR}).
The instances used are the networks that we extracted for the scalability experiments.
We measure the time for computing a maximum preflow.
In case of \Partitioner{Dinitz} algorithm the measurement also includes the time for deriving the source-side cut.
The parallel algorithms are run with one thread, in order to investigate the overheads of using parallel data structures and loop constructs.
As we can see, the graph-based push-relabel implementation (\Partitioner{LawlerPR}) is consistently a factor of two or more slower.
Note that all push-relabel versions use the additional capacities optimization from Section~\ref{sec:flows:impl-details}.
Using parallel queues for active nodes and global relabeling in the sequential FIFO code (\Partitioner{SeqPR-ParQueue}) is consistently slower, but not by much.
In previous versions of the framework~\cite{REBAHFC, KAHYPAR-HFC}, \Partitioner{Dinitz} algorithm~\cite{Dinitz} was used because it is well suited for the incremental flow problems of \Partitioner{FlowCutter}.
This may have been an oversight, as \Partitioner{Dinitz} performs worse than \Partitioner{SeqPR} on roughly $\frac{2}{3}$ of the instances, often by a substantial margin; though this measurement does not consider incremental flow problems.
As opposed to the push-relabel versions, \Partitioner{SeqPR} does not strictly outperform \Partitioner{Dinitz}, since \Partitioner{Dinitz} is noticeably faster on $\frac{1}{3}$ of the instances.
We consider two parallel push-relabel implementations that differ in the way they address the relabeling bug (confer Section~\ref{sec:flows:bug}): run global relabeling before termination (\Partitioner{ParPR-RL}) or block nodes from relabeling after being pushed to (\Partitioner{ParPR-Block}).
They perform completely equivalently.
We observe very moderate slowdowns over \Partitioner{SeqPR}, ranging mostly between 1 and 2.

\end{document}